\documentclass[trackchanges]{aastex701}

\usepackage{amsmath} 
\usepackage{amssymb}
\usepackage{graphicx}
\usepackage{dirtytalk}
\usepackage{natbib}
\usepackage{units}

\newcommand{\iu}{\mathrm{i}\mkern1mu}

\newcommand{\ddt}[1]{\frac{{\rm{d}}{#1}}{{\rm{d}}t}}
\newcommand{\prp}[1]{\frac{\partial\mathcal{R}}{\partial {#1}}}

\begin{document}

\title{Analytical formulae for planet-planet or moon-moon tides}

\author[orcid=0000-0002-6779-3848]{Michaela Walterová}
\affiliation{Department of Geophysics, Faculty of Mathematics and Physics, Charles University, Prague, Czech Republic}
\email[show]{Michaela.Walterova@matfyz.cuni.cz}  

\author[orcid=0000-0001-8227-0685]{Marie Běhounková}
\affiliation{Department of Geophysics, Faculty of Mathematics and Physics, Charles University, Prague, Czech Republic} 
\email{Marie.Behounkova@matfyz.cuni.cz}

\correspondingauthor{Michaela Walterová}


\begin{abstract}

Almost negligible among the solar system planets, mutual tides raised by a secondary on another secondary orbiting the same primary may gain increased importance in tightly packed systems, e.g., between the components of close-in exoplanetary systems or between the moons of gas giants. In this work, we derive an analytical formula for the planet-planet tidal potential and for the effect of mutual tides on the rotational and orbital evolution of moons or planets in a coplanar configuration. The procedure follows closely the steps of W. M. Kaula (1961), and the resulting potential and derived quantities are presented in the form of an expansion into modes of different frequencies (in the temporal domain) and spherical harmonic components (in the spatial domain). The analytical formulae are implemented into a code and compared with an independent numerical solution.

\end{abstract}



\section{Introduction} 

Tidal interaction is one of the major phenomena influencing the long-term orbital and thermal evolution of close-in exoplanets and planetary satellites. Apart from the tidal deformation and heating of planetary (or lunar) interiors, the mutual gravitational action between the primary and the secondary, both considered as extent bodies, results in the variations of orbital elements and of the spin rate. The expected final state of tidally evolved close-in rocky exoplanets or moons without an atmosphere is then a circular orbit and synchronous rotation \citep[e.g.,][]{correia14}. 

The orbital dynamics of multiplanetary systems are also affected by the gravitational interaction between secondaries. In this context, the secondaries are typically treated as point masses and their effect on the orbits of each other is described as a third-body perturbation \citep{murray99}. However, in tightly packed systems, the differences in the gravitational force acting throughout an extent (i.e., not point-mass) secondary due to another secondary might not be negligible. In such cases, where the tidal deformation is not caused by the primary, but rather by another secondary, we can speak of planet--planet or moon--moon tides. 

To our knowledge, the first discussion of planet--planet tides was provided by \citet{lingam18}, who emphasised the astrobiological significance of this phenomenon. According to their estimates, the tidal elevation on the TRAPPIST-1 exoplanets caused by planet--planet tides should be comparable to the elevation caused on the Earth by the tidal action of the Moon. Moreover, since the TRAPPIST-1 system constitutes a resonant chain, this effect can be further enhanced during multiple conjunctions. If the surfaces of the planets are partially covered by an ocean, the complex tidal modulation caused by the conjunctions may have important consequences for abiogenesis and evolution of life \citep{lingam18}. Later that year, a research note dedicated to the planet--planet tides in the TRAPPIST-1 system was published by \citet{wright18b}. \citet{wright18b} estimated that the tidal strain and tidal heat rate produced in planet $g$ due to the action of planet $f$ may be even greater than the deformation and dissipation induced by the tidal interaction with the host star. For the other planets (except $b$), the ratio between the strain due to planet--planet tides and the star--planet tides was calculated as being higher than $10\%$; the same ratio for the tidal dissipation being above $1\%$. Although these numbers were presented as first-order estimates, they clearly indicate that planet-planet tidal loading should be included in detailed studies of the system.

A more complete examination of the TRAPPIST-1 planet-planet tides was subsequently conducted by \citet{hay19}. The authors discussed the role of viscosity and orbital eccentricity on the relative effect of planet--planet tides (compared to the star--planet tides) and reported lower values for their contribution to the overall heating than estimated by \citet{wright18b}. Nevertheless, for planet $g$, they still predicted that planet--planet tidal heating may participate by up to $20\%$ on the total tidal heating. In a later study, focused on the Galilean satellites, \citet{hay20} suggested that planet--planet (or moon--moon) tides may be particularly important in ocean worlds, where they can excite resonant response and thus alter the system's dynamics.\\

The main purpose of this work is to provide an analytical formula for the tidal potential due to planet--planet (or moon--moon) loading, transformed to a planetocentric coordinate system and expressed by means of Keplerian elements. Along with a series expansion of the tidal potential, we provide a general expression for all tidal frequencies involved in planet-planet loading and several other quantities. Derivation of the tidal potential (Section \ref{sec:pl-pl-derivation}) follows closely the steps of \citet{kaula61} and can be understood as Fourier decomposition. In Section \ref{sec:pl-pl-torque}, we derive the tidal torque resulting from planet--planet tides and potentially contributing to the secular rotational evolution and physical libration of the planets. Section \ref{sec:evol} complements the tidal torque with orbital evolution equations. Discussion and testing of the analytical formulae against an independent numerical implementation (introduced in Appendix \ref{sec:appA}) is provided in Section \ref{sec:comparison}. The work is then concluded by Section \ref{sec:conclusion}. Since the derivation is algebraically heavy, we limit the main text to a simplified, coplanar case and we leave the full 3-D case for Appendix \ref{sec:supplement}.

\section{Derivation of the tidal potential} \label{sec:pl-pl-derivation}

\begin{figure}
    \centering
    \includegraphics[width=0.4\textwidth]{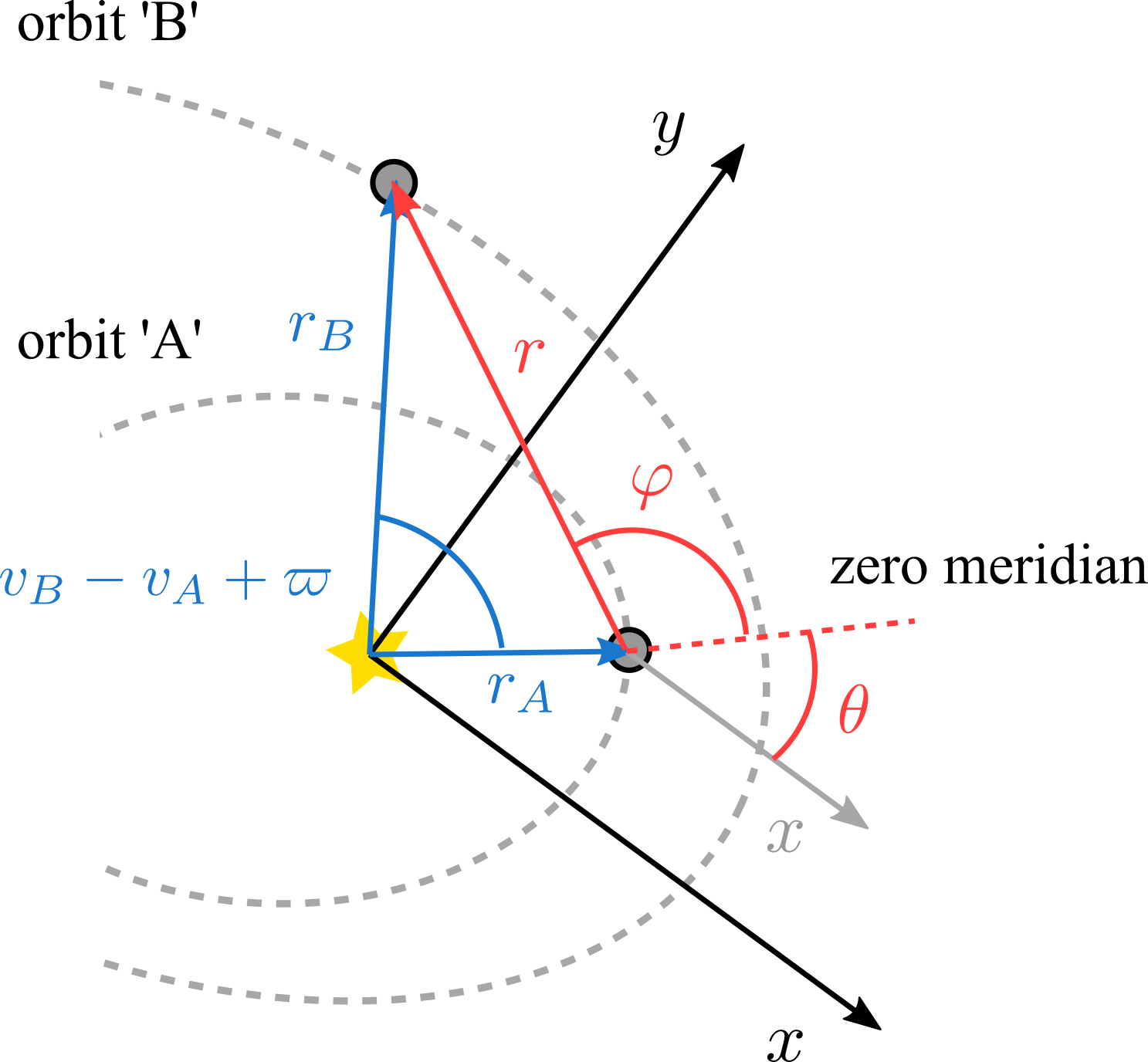}
    \vspace*{5mm}
    \caption{The geometry of a biplanetary system and the introduction of several quantities required by the derivation: the position vectors $r_{\rm{A}}$ and $r_{\rm{B}}$ of planets $A$ and $B$, respectively, the angle $v_B-v_A+\varpi$ between the two position vectors, the mutual planet-planet distance $r$, the sidereal time $\theta$, and the planetocentric longitude $\varphi$ of planet $B$, measured from planet $A$'s zero meridian.}
    \label{fig:geometry}
\end{figure}

Consider a system consisting of a primary and two secondaries. The primary can be a host star or a gas giant, the secondaries can be planets or moons, respectively. For simplicity, we will call the primary ``a star" and the secondaries will be denoted as ``planet $A$" and ``planet $B$" (Figure \ref{fig:geometry}). Each of the planets orbits the star on a Keplerian trajectory described by a set of six elements $\{a_k,e_k,i_k,\omega_k,\Omega_k,\sigma_k\}$: the semi--major axis, the orbital eccentricity, the inclination of the orbit to a reference plane, the argument of the periapsis, the longitude of ascending node and the mean anomaly at epoch, respectively. Let us also assume that the orbits of the two planets are coplanar and the obliquities of all bodies in the system are equal to zero\footnote{A more general, yet extremely complex formula for the fully 3-D case can be found in Appendix \ref{sec:supplement}.}. This simplified case may be still meaningful because compact exoplanetary systems, as well as regular satellites of solar system gas giants, present only a minor deviation from coplanarity \citep[e.g.,][]{jacobson2022,knudstrup2024}. Additionally, gravitational tidal interaction tends to align the planets' or moons' spin axes with the orbital plane, unless balanced by atmospheric tides \citep{boue16,valente2024}.

Let $A$ be the planet of our interest, whose deformation is to be studied, and let $B$ be the perturber. The instantaneous position of the two bodies in an asterocentric reference frame $(x,y)$ is described by the radial distance from the host star, $r_A$ and $r_B$, and by the true anomaly, $v_A$ and $v_B$. The angle between position vectors of their periapses---or the relative longitude of $B$'s pericenter---will be denoted by $\varpi=\varpi_A-\varpi_B=(\omega_A+\Omega_A)-(\omega_B+\Omega_B)$. Our goal is to express the external potential sensed at any point of $A$ due to the presence of $B$ as a function of the orbital elements $a_A$, $a_B$, $e_A$, $e_B$, $\varpi_A$, $\varpi_B$, the mean anomalies $M_A=n_A t + \sigma_A$ and $M_B=n_B t + \sigma_B$, and time $t$. Symbols $n_A$ and $n_B$ stand for the mean motions of the two planets.\\

We begin with expressing the external potential acting on infinitesimal volume of planet $A$ in the form of a linear combination of spherical harmonics \citep{kaula61}

\begin{equation}
\mathcal{U} = \sum_{l=0}^{\infty} \frac{1}{r^{l+1}}\; \sum_{m=0}^{l}\mathcal{P}_{lm}(\cos\vartheta)\, \Big(\mathcal{A}_{lm}\cos{m\varphi}+\mathcal{B}_{lm}\sin{m\varphi}\Big)\; ,
\label{eq:potentialsimpl}
\end{equation}
where $(r,\vartheta,\varphi)$ are planetocentric equatorial coordinates of the perturber referred to the center of planet $A$, functions $\mathcal{P}_{lm}(\cos\vartheta)$ are the associated Legendre polynomials of degree $l$ and order $m$, defined by the following formula \citep[e.g.,][]{kaula61},\\

\begin{equation}
\mathcal{P}_{lm}(x) = \frac{(1-x^2)^{\frac{m}{2}}}{2^l l!}\; \sum_{t=0}^{\frac{l-m}{2}} \frac{(2l-2t)!}{(l-m-2t)!} \binom{l}{t} (-1)^t x^{l-m-2t}\; , \label{eq:legendresimpl}
\end{equation}
and $\mathcal{A}_{lm}$ and $\mathcal{B}_{lm}$ stand for a combination of physical parameters and algebraic terms arising from the formula for gravitational potential and from the addition theorem,

\begin{align*}
\mathcal{A}_{lm} &= (2-\delta_{m0})\frac{(l-m)!}{(l+m)!}\; \mathcal{G}\mathcal{M}_{B}(r')^{l}\; \mathcal{P}_{lm}(\cos\vartheta')\; \cos{m\varphi'}\; , \\[1.5ex]
\mathcal{B}_{lm} &= (2-\delta_{m0})\frac{(l-m)!}{(l+m)!}\; \mathcal{G}\mathcal{M}_{B}(r')^{l}\; \mathcal{P}_{lm}(\cos\vartheta')\; \sin{m\varphi'}\; .
\end{align*}\\
In the last two expressions, $\mathcal{G}$ is Newton's gravitational constant, $\mathcal{M}_B$ mass of the perturber and $(r',\vartheta',\varphi')$ coordinates of the ``observer" (i.e., of the perturbed infinitesimal volume).\\

The tidal potential introduced by equation (\ref{eq:potentialsimpl}) depends on the planetographic position of the perturber through the associated Legendre polynomials and the goniometric functions $\cos{m\varphi}$ and $\sin{m\varphi}$. From the geometry of the system, we can express the longitude of the point at planet $A$'s surface above which the perturber culminates (the ``sub-perturber point") as

\begin{equation}
    \varphi = \alpha - \theta,
\end{equation}\\
where $\alpha$ is the right ascension of planet $B$ and $\theta$ is the sidereal time (or the rotation angle of planet $A$). Then, we have

\begin{align*}
    \cos{m\varphi} &= \cos{m\alpha}\cos{m\theta} + \sin{m\alpha}\sin{m\theta}\; , \\[1.5ex]
    \sin{m\varphi} &= \sin{m\alpha}\cos{m\theta} - \cos{m\alpha}\sin{m\theta}\; ,
\end{align*}\\
and with the help of trigonometric identities (3) and (4) of \citet{kaula61},

\begin{equation}
\begin{aligned}
\cos{mx} &= \Re\Bigg\{\sum_{s=0}^{m}\binom{m}{s} \iu^{s} \cos^{m-s}x \sin^{s}x \Bigg\}\; , \\[1.5ex]
\sin{mx} &= \Re\Bigg\{\sum_{s=0}^{m}\binom{m}{s} \iu^{s-1} \cos^{m-s}x \sin^{s}x \Bigg\}\; ,
\end{aligned}
\end{equation}\\
we can rewrite the combination of goniometric functions in (\ref{eq:potentialsimpl}) as

\begin{align}
\mathcal{A}_{lm}\cos{m\varphi}+\mathcal{B}_{lm}\sin{m\varphi} =\Re\Bigg\{\sum_{s=0}^{m}\binom{m}{s} \iu^{s} \cos^{m-s}\alpha\; \sin^{s}\alpha\; \big(\mathcal{A}_{lm}-\iu\mathcal{B}_{lm}\big) \exp\big[-\iu\,m\,\theta\big] \Bigg\}\, ,
\label{eq:abcossin1}
\end{align}\\
where $\Re$ symbolizes the real part. As a consequence of the coplanarity of both orbits and the zero obliquity of $A$'s spin axis, planet $B$ is always located above a point with colatitude $\vartheta=\frac{\pi}{2}$, while the longitude of sub--$B$ point depends only on the two true anomalies and the instantaneous mutual distances between the three bodies (two planets and the star) in the system. The functions $\cos{\alpha}$ and $\sin{\alpha}$, multiplied by the instantaneous mutual distance of the two planets $r$, are the $x$-- and $y$--components of the position vector $\mathbf{r} = \mathbf{r}_B - \mathbf{r}_A$, which defines the position of planet $B$ with respect to planet $A$. It follows that
\begin{equation}
\begin{gathered}
\cos{\alpha} = \frac{r_A\cos{(v_A+\varpi_A)}-r_B\cos(v_B+\varpi_B)}{r}\, , \\[1.5ex]
\sin{\alpha} = \frac{r_A\sin{(v_A+\varpi_A)}-r_B\sin(v_B+\varpi_B)}{r}\, .
\label{eq:goniosimpl}
\end{gathered}
\end{equation}\\
and
\begin{equation}
r = \sqrt{r_{A}^2+r_{B}^2-2\,r_{A}\,r_{B}\,\cos(v_B-v_A+\varpi)}\; .
\label{eq:mutdistsimpl}
\end{equation}\\
The only non--zero terms in the explicit definition of associated Legendre polynomials (\ref{eq:legendresimpl}) for $\vartheta=\frac{\pi}{2}$ are those with $t=\frac{l-m}{2}$, where $t$ is an integer. This condition implies that polynomials with odd $(l-m)$ are identically zero,

\begin{equation}
\mathcal{P}_{lm}(\cos\vartheta=0) = 0 \qquad \iff \qquad (l-m) \quad \mbox{odd}\; ,
\end{equation}\\
while polynomials with even $(l-m)$ simplify to

\begin{equation}
\mathcal{P}_{lm}(\cos\vartheta=0) = \frac{1}{2^l}\; \frac{(l+m)!}{l!}\; \binom{l}{\frac{l-m}{2}} \; (-1)^{\frac{l-m}{2}}\qquad \iff \qquad (l-m) \quad \mbox{even}\; . \label{eq:legendrecossimpl}
\end{equation}\\

In the next step, we further expand equation (\ref{eq:abcossin1}), which includes expressions $\cos^{m-s}{\alpha}$ and $\sin^{s}{\alpha}$. Applying binomial theorem on equations (\ref{eq:goniosimpl}), we obtain

\begin{equation}
\begin{gathered}
\cos^{m-s}{\alpha} = \frac{1}{r^{m-s}}\; \sum_{c_1=0}^{m-s} (-1)^{c_1} \binom{m-s}{c_1}\; r_A^{m-s-c_1}\; r_B^{c_1}\; \cos^{m-s-c_1}{(v_A+\varpi_A)}\;\cos^{c_1}{(v_B+\varpi_B)}\; , \\[1.5ex]
\sin^{s}{\alpha} = \frac{1}{r^{s}}\; \sum_{d_1=0}^{s} (-1)^{d_1} \binom{s}{d_1}\; r_A^{s-d_1}\; r_B^{d_1} \; \sin^{s-d_1}{(v_A+\varpi_A)}\;\sin^{d_1}{(v_B+\varpi_B)}\; .
\end{gathered} \label{eq:binom}
\end{equation}\\
We see that the product of the two terms expanded in (\ref{eq:binom}) contains products of goniometric functions with the same arguments, namely $(v_A+\varpi_A)$ or $(v_B+\varpi_B)$. This kind of expressions can be rewritten with the help of another goniometric identity, namely equation (5) of \citet{kaula61}:

\begin{equation}
\sin^a{x} \cos^b{x} = \frac{(-\iu)^a}{2^{a+b}} \sum_{c=0}^{a} \sum_{d=0}^{b} \binom{a}{c}\binom{b}{d} (-1)^c \big[\cos(a+b-2c-2d)x+\iu\sin(a+b-2c-2d)x\big]\; . \label{eq:sinacosbsimpl}
\end{equation}\\
The resulting products of goniometric functions are

\begin{align}
&\cos^{c_1}{(v_B+\varpi_B)}\; \sin^{d_1}{(v_B+\varpi_B)} = \frac{(-\iu)^{d_1}}{2^{c_1+d_1}}\; \sum_{h_1=0}^{d_1} \sum_{h_2=0}^{c_1} \binom{d_1}{h_1}\; \binom{c_1}{h_2}\;\nonumber\times\\[2ex]
& \hspace{3.5cm}\times \; (-1)^{h_1}\; \exp\Big[\iu\,(c_1+d_1-2h_1-2h_2)\,(v_B+\varpi_B)\Big] \nonumber\\[2ex]
&\cos^{m-s-c_1}{(v_A+\varpi_A)}\; \sin^{s-d_1}{(v_A+\varpi_A)} = \\[2ex]
&\hspace{3.5cm} =\;\frac{(-\iu)^{s-d_1}}{2^{m-c_1-d_1}}\; \sum_{k_1=0}^{s-d_1} \sum_{k_2=0}^{m-s-c_1} \binom{s-d_1}{k_1}\; \binom{m-s-c_1}{k_2}\;\times\nonumber\\[2ex]
&\hspace{3.5cm}\times\; (-1)^{k_1}\; \exp\Big[\iu\,(m-c_1-d_1-2k_1-2k_2)\,(v_A+\varpi_A)\Big] \nonumber
\end{align}\\
Inserting both of these expansions into (\ref{eq:abcossin1}) and consequently into (\ref{eq:potentialsimpl}) leads us to a preliminary form of the tidal potential,

\overfullrule=0pt
\begin{equation}
\begin{gathered}
\mathcal{U}_{lm} = \frac{1}{r^{l+m+1}}\; \frac{(l+m)!}{2^{l+m}\, l!}\; \binom{l}{\frac{l-m}{2}}\; (-1)^{\frac{l-m}{2}}\; \sum_{s=0}^{m} \binom{m}{s}\; \sum_{c_1=0}^{m-s} \sum_{d_1=0}^{s} \binom{m-s}{c_1}\; \binom{s}{d_1}\; r_B^{c_1+d_1}\; r_A^{m-c_1-d_1}\;  \times \\[2ex]
\times\;(-1)^{c_1+d_1}\; \sum_{h_1=0}^{d_1} \sum_{h_2=0}^{c_1} \binom{d_1}{h_1}\; \binom{c_1}{h_2}\; (-1)^{h_1}\; \sum_{k_1=0}^{s-d_1} \sum_{k_2=0}^{m-s-c_1} \binom{s-d_1}{k_1}\; \binom{m-s-c_1}{k_2}\; (-1)^{k_1}\; \times \\[2ex]
\times\; \Re\Bigg\{\Big(\mathcal{A}_{lm}-\iu\mathcal{B}_{lm}\Big)\; \exp\Big\{\iu\Big[(c_1+d_1-2h_1-2h_2)\,(v_B+\varpi_B) + (m-c_1-d_1-2k_1-2k_2)\,(v_A+\varpi_A) - m\theta\Big]\Big\}\Bigg\}\; .
\end{gathered} \label{eq:usimpl}
\end{equation}\\
\overfullrule=5pt
This form can be further simplified by introducing a new set of indices and deriving their lower and upper limits:

\begin{equation}
\begin{aligned}
c_1 &= p_1 - d_1\; , \qquad p_1 \in [0,m]\;, \\[2ex]
h_2 &= p_2 - h_1\; , \qquad p_2 \in [0,p_1]\;, \\[2ex]
k_2 &= p_3 - k_1\; , \qquad p_3 \in [0,m-p_1]\;.
\end{aligned}
\end{equation}\\
Together with setting the limits for the new indices $p_1$, $p_2$ and $p_3$, we should also update the limits for the old indices $d_1$, $h_1$ and $k_1$ because some combinations of the new and the old indices result in zero binomial coefficients in (\ref{eq:usimpl}). Updated intervals of the values of old indices are:

\begin{equation}
\begin{aligned}
\max(0,p_1-m+s) \;\leq\; &d_1 \;\leq\; \min(s,p_1)\;, \\[2ex]
\max(0,p_2-p_1+d_1) \;\leq\; &h_1 \;\leq\; \min(d_1,p_2) \;,\\[2ex]
\max(0,p_3+p_1-m+s-d_1) \;\leq\; &k_1 \;\leq\; \min(s-d_1,p_3)\;.
\end{aligned} \label{eq:lims1}
\end{equation}\\

Up to now, we have succeeded in rewriting the planetographic longitude and latitude of the sub--perturber point in terms of the desired variables related to the instantaneous position of both planets and the rotation angle $\theta$ of planet $A$. The last term to be transformed to these elements is the mutual distance between the two bodies, $r$, introduced by equation (\ref{eq:mutdistsimpl}). Specifically, we are seeking a series expansion of the function $r^{-\lambda}$ (for $\lambda=l+m+1$), which is equivalent to the generating function of Gegenbauer polynomials \citep[e.g.,][]{reimer03},

\begin{equation}
\frac{1}{\left(\sqrt{1+t^2-2xt}\right)^{\lambda}} = \sum_{n=0}^{\infty} t^n \, \mathcal{C}_{n}^{\left(\frac{\lambda}{2}\right)}(x) \label{eq:gegendef}
\end{equation}\\
with $\lambda\in\mathbb{R}$, $\lambda>-1$ and $\lambda\neq0$. The Gegenbauer polynomials $\mathcal{C}_{n}^{\left(\frac{\lambda}{2}\right)}(x)$ are defined as

\begin{equation}
\mathcal{C}_{n}^{\left(\frac{\lambda}{2}\right)}(x) = \sum_{u=0}^{\lfloor\frac{n}{2}\rfloor} (-1)^{u}\, \frac{\Gamma\left(n-u+\frac{\lambda}{2}\right)}{\Gamma\left(\frac{\lambda}{2}\right)\,u!(n-2u)!}\,(2x)^{n-2u}\; , \label{eq:gegensimpl}
\end{equation}\\
where $\lfloor\frac{n}{2}\rfloor$ denotes the integer part of $\frac{n}{2}$. Applying expansions (\ref{eq:gegendef}) and (\ref{eq:gegensimpl}) to our case, we obtain the following expression:

\begin{equation}
\frac{1}{r^{l+m+1}} = \frac{1}{r_{>}^{l+m+1}}\; \sum_{n=0}^{\infty} \left(\frac{r_{<}}{r_{>}}\right)^n\; \sum_{u=0}^{\lfloor\frac{n}{2}\rfloor} (-1)^{u}\, \frac{\Gamma\left(n-u+\frac{l+m+1}{2}\right)}{\Gamma\left(\frac{l+m+1}{2}\right)\,u!(n-2u)!}\,\big[2\cos(v_B-v_A+\varpi)\big]^{n-2u}\; , \label{eq:rexpsimpl}
\end{equation}\\
where $r_{>}$ and $r_{<}$ indicate the longer and the shorter one of the instantaneous radial distances of each planet from the host star. Once again, we apply the trigonometric identity (\ref{eq:sinacosbsimpl}) on the function $\big[\cos(v_B-v_A+\varpi)\big]^{n-2u}$ and obtain the final expansion of (\ref{eq:rexpsimpl}):

\begin{equation}
\begin{aligned}
\frac{1}{r^{l+m+1}} = \frac{1}{r_{>}^{l+m+1}}\; \sum_{n=0}^{\infty} \left(\frac{r_{<}}{r_{>}}\right)^n\; &\sum_{u=0}^{\lfloor\frac{n}{2}\rfloor} (-1)^{u}\, \frac{\Gamma\left(n-u+\frac{l+m+1}{2}\right)}{\Gamma\left(\frac{l+m+1}{2}\right)\,u!(n-2u)!}\; \times \\[2ex]
&\times\; \sum_{j=0}^{n-2u} \binom{n-2u}{j}\, \exp\Big[\iu\,(n-2u-2j)(v_B-v_A+\varpi)\Big] \; .
\end{aligned} \label{eq:rfinsimpl}
\end{equation}\\
Similarly to the previous expansion, it is convenient to introduce a new index, $p_4$, and to update the limits for the summation over $u$:

\begin{equation}
\begin{gathered}
j = p_4 - u\; , \qquad p_4 \in [0,n]\;, \\[2ex]
0 \;\leq\; u \;\leq\; \min\left(p_4,n-p_4\right)\;. \\[2ex]
\end{gathered} \label{eq:lims2}
\end{equation}\\
Finally, we insert (\ref{eq:rfinsimpl}) into the preliminary form of the tidal potential (\ref{eq:usimpl}) and gather all algebraic prefactors into an analogy of Kaula's functions of inclination, $\mathcal{F}^0_{lmn\bar{p}}$, where the superscript $0$ indicates that we are in the limit of coplanar orbits and zero obliquities and $\bar{p}$ stands for the multiindex $\{p_1,p_2,p_3,p_4\}$. The tidal potential can then be rewritten as

\overfullrule=0pt
\begin{equation}
\begin{gathered}
\mathcal{U}_{lm} = \sum_{n=0}^{\infty} \sum_{p_1=0}^{m} \sum_{p_2=0}^{p_1} \sum_{p_3=0}^{m-p_1} \sum_{p_4=0}^{n} \frac{r_B^{p_1}\, r_A^{m-p_1}}{r_{>}^{l+m+1}}\; \left(\frac{r_{<}}{r_{>}}\right)^n\; \mathcal{F}^0_{lmn\bar{p}}\; \times \\[2ex]
\times\; \Bigg\{
\mathcal{A}_{lm}
\cos\Big[(m-p_1-2p_3-n+2p_4)(v_A+\varpi_A) +(p_1-2p_2+n-2p_4)(v_B+\varpi_B)- m\theta\Big] \; + \\[2ex]
+\;
\mathcal{B}_{lm}
\sin\Big[(m-p_1-2p_3-n+2p_4)(v_A+\varpi_A) + (p_1-2p_2+n-2p_4)(v_B+\varpi_B) - m\theta\Big]\Bigg\}
\end{gathered} \label{eq:potcirc}
\end{equation}\\
\overfullrule=5pt
with

\begin{equation}
\begin{gathered}
\mathcal{F}^0_{lmn\bar{p}} = \frac{1}{2^{l+m}}\; \frac{(l+m)!}{l!}\; \binom{l}{\frac{l-m}{2}} (-1)^{\frac{l-m}{2}} \sum_{s=0}^{m} \binom{m}{s} \sum_{d_1} \binom{m-s}{p_1-d_1} \binom{s}{d_1} (-1)^{p_1} \; \times \\[2ex]
\times\; \sum_{h_1} \sum_{k_1} (-1)^{h_1+k_1} \binom{d_1}{h_1} \binom{p_1-d_1}{p_2-h_1} \binom{s-d_1}{k_1} \binom{m-s-p_1+d_1}{p_3-k_1} \; \times \\[2ex]
\times\; \sum_{u} (-1)^{u} \frac{\Gamma\left(n-u+\frac{l+m+1}{2}\right)}{\Gamma\left(\frac{l+m+1}{2}\right)\,u!\,(n-2u)!}\; \binom{n-2u}{p_4-u}
\end{gathered} \label{eq:fpref}
\end{equation}\\
and limits of summation over $d_1$, $h_1$, $k_1$ and $u$ given by (\ref{eq:lims1}) and (\ref{eq:lims2}), respectively. We also reiterate that the above expression holds only for even $(l-m)$, as the terms $\mathcal{U}_{lm}$ for odd $(l-m)$ equal zero.\newline

Equation (\ref{eq:potcirc}) can be readily used for the computation of planet--planet tidal potential in the case of circular orbits. For generally eccentric orbits, we need to expand the instantaneous distances of both planets from the host star and the true anomalies into series containing the respective semi--major axes $a_i$ and the mean anomalies $M_i$. Let $X_{k}^{b,c}$ denote Hansen's coefficients \citep[e.g.,][]{murray99}. Then,

\begin{equation}
\left(\frac{r_i}{a_i}\right)^b e^{icv_i} = \sum_{k=-\infty}^{\infty} X_{k}^{b,c}(e_i) e^{ikM_i} \; . \label{eq:hansensimpl}
\end{equation}\\
Here, Hansen's coefficients play the role of Kaula's short-period functions of eccentricity $\mathcal{G}_{lpq}(e)$, as defined in \citet{kaula61}. In order to assign correct values to the exponent $b$, it is necessary to distinguish two particular cases. The instantaneous distance of planet $B$ from the host star is either always larger than the distance of planet $A$,

\begin{equation}
a_B > a_A \frac{1+e_A}{1-e_B}\; ,
\end{equation}\\
or it is always smaller, hence
\begin{equation}
a_A > a_B \frac{1+e_B}{1-e_A}\; .
\end{equation}\\
Other configurations would result in crossing orbits, a case that is not considered in our derivation.

If $B$ is the outer planet, we may substitute

\begin{equation*}
b_A = m+n-p_1\; , \qquad \qquad b_B = -m-n+p_1-l-1\; ,
\end{equation*}
otherwise, if $B$'s orbit lies inside $A$'s orbit, we have

\begin{equation*}
b_A = -n-p_1-l-1\; , \qquad \qquad b_B = n+p_1\;.
\end{equation*}
The other two indices are

\begin{align*}
c_A &= m-p_1-2p_3-n+2p_4 && c_B = p_1-2p_2+n-2p_4\\
k_A &= m-p_1-2p_3-n+2p_4+q_1 && k_B = p_1-2p_2+n-2p_4+q_2\;.
\end{align*}\\
Introducing functions of eccentricity $\mathcal{G}_{lmn\bar{p}\bar{q}}^A(e_A)=X_{k_A}^{b_A,c_A}(e_A)$ and $\mathcal{G}_{lmn\bar{p}\bar{q}}^B(e_B)=X_{k_B}^{b_B,c_B}(e_B)$, where $\bar{q}$ is the multiindex $\{q_1, q_2\}$, we may write down the final form of the tidal potential due to planet--planet tides as

\begin{equation}
\begin{gathered}
\mathcal{U}_{lm} = \sum_{n=0}^{\infty} \sum_{p_1=0}^{m} \sum_{p_2=0}^{p_1} \sum_{p_3=0}^{m-p_1} \sum_{p_4=0}^{n} \sum_{q_1=-\infty}^{\infty} \sum_{q_2=-\infty}^{\infty} \frac{1}{a_{>}^{l+1}}\; \left(\frac{a_{<}}{a_{>}}\right)^{\varrho}\; \mathcal{F}^0_{lmn\bar{p}}\; \mathcal{G}_{lmn\bar{p}\bar{q}}^A(e_A)\; \mathcal{G}_{lmn\bar{p}\bar{q}}^B(e_B) \times \\[2ex]
\times\; \Big\{\mathcal{A}_{lm}
\cos\Big[(m-p_1-2p_3-n+2p_4+q_1)M_A +(p_1-2p_2+n-2p_4+q_2)M_B + \\[2ex]
+ (m-p_1-2p_3-n+2p_4)\varpi_A +(p_1-2p_2+n-2p_4)\varpi_B - m\theta\Big] \; + \\[2ex]
+\;
\mathcal{B}_{lm}
\sin\Big[(m-p_1-2p_3-n+2p_4+q_1)M_A + (p_1-2p_2+n-2p_4+q_2)M_B + \\[2ex]
+ (m-p_1-2p_3-n+2p_4)\varpi_A + (p_1-2p_2+n-2p_4)\varpi_B - m\theta\Big]\Big\}
\end{gathered} \label{eq:potecc}
\end{equation}\\
with the algebraic prefactors $\mathcal{F}^0_{lmn\bar{p}}$ given by (\ref{eq:fpref}). The exponent $\varrho$ depends on the geometry of the two orbits. If $A$ is the inner planet, perturbed by outer $B$, then $\varrho=m+n-p_1$. In the opposite case of outer $A$, we have $\varrho=n+p_1$. For odd $l-m$, the term $\mathcal{U}_{lm}$ is identically zero.

The argument of function (\ref{eq:potecc}) has essentially the form $\omega_{mn\bar{p}\bar{q}}t+\gamma_0$, where $\omega_{mn\bar{p}\bar{q}}$ is the signed frequency of a given tidal mode and $\gamma_0$ the initial phase:

\begin{equation}
    \begin{gathered}
    \omega_{mn\bar{p}\bar{q}}=(m-p_1-2p_3-n+2p_4+q_1)n_A +(p_1-2p_2+n-2p_4+q_2)n_B + \\[0.5em]
    + (m-p_1-2p_3-n+2p_4)\dot{\varpi}_A +(p_1-2p_2+n-2p_4)\dot{\varpi}_B - m\dot{\theta}
    \end{gathered}
\end{equation}\\
and

\begin{equation}
    \begin{gathered}
    \gamma_0=(m-p_1-2p_3-n+2p_4+q_1)\sigma_A +(p_1-2p_2+n-2p_4+q_2)\sigma_B + \\[0.5em]
    + (m-p_1-2p_3-n+2p_4){\varpi}_{A0} +(p_1-2p_2+n-2p_4){\varpi}_{B0} - m\theta_0\; .
    \end{gathered}
\end{equation}\\
%

\section{Tidal torque} \label{sec:pl-pl-torque}
Equation (\ref{eq:potecc}) presents harmonic expansion of the tidal potential acting on infinitesimal volume of planet $A$ due to the presence of planet $B$. As a consequence of nonhomogeneity of the tidal potential, planet $A$ deforms and its departure from a perfectly spherical shape gives rise to disturbing potential $\delta\mathcal{U}$, whose value exterior to the body reads as \citep{kaula64}

\begin{equation}
\begin{aligned}
\delta\mathcal{U}_{lmn\bar{p}\bar{q}}^{B}(\mathbf{r}') &= \left(\frac{R_{A}}{r'}\right)^{l+1}\, k_{lm}(\omega_{mn\bar{p}\bar{q}}) \Big[\mathcal{U}_{lmn\bar{p}\bar{q}}(R_{A})\Big]_{{\rm{lag}}} \\[2ex]
&\hspace{-1.2cm}= k_{lm}(\omega_{mn\bar{p}\bar{q}})\, \frac{R_{A}^{2l+1}}{(r')^{l+1}} \, B_{lm}^{B} \, C_{lmn\bar{p}\bar{q}}^{B} \, \mathcal{P}_{lm}(\cos{\vartheta'})
\cos
\Big[\nu_{mn\bar{p}\bar{q}}-\varepsilon_{lm}(\omega_{mn\bar{p}\bar{q}}) - m(\varphi'+\theta)\Big]\; .
\end{aligned} \label{eq:disturbing}
\end{equation}\\
The subscript ``lag" reminds us that, in order to compute the disturbing potential of non--elastic body at time $t$, we need to express the tidal potential at time $t-\Delta t_{lmn\bar{p}\bar{q}}$, where $\Delta t_{lmn\bar{p}\bar{q}} = \varepsilon_{lm}(\omega_{mn\bar{p}\bar{q}})/\omega_{mn\bar{p}\bar{q}}$ is a mode--specific tidal time lag. Parameters $k_{lm}$ and $\varepsilon_{lm}$ are, respectively, the tidal Love number and the tidal phase lag: two quantities, whose magnitude and frequency dependence are given by the internal structure and the rheology of the planet. For a radially symmetric body, both $k_{lm}$ and $\varepsilon_{lm}$ are independent of the order $m$ and we can write simply $k_l$ and $\varepsilon_{l}$.

In (\ref{eq:disturbing}), we also introduced the traditional notation for the non--rotational part of the argument,

\begin{equation}
\begin{aligned}
\nu_{mn\bar{p}\bar{q}} &= (m-p_1-2p_3-n+2p_4+q_1)\,M_A  + (p_1-2p_2+n-2p_4+q_2)\,M_B + \\[2ex]
&+ (m-p_1-2p_3-n+2p_4)\,\varpi_A +  (p_1-2p_2+n-2p_4)\,\varpi_B \; ,
\end{aligned}
\end{equation}
and the two prefactors
\begin{equation}
B_{lm}^{B} = \mathcal{G}\mathcal{M}_{B}\, \frac{(l-m)!}{(l+m)!}\, (2-\delta_{m0})\; ,
\end{equation}\\
\begin{equation}
C_{lmn\bar{p}\bar{q}}^{B} = \frac{1}{a_{>}^{l+1}}\; \left(\frac{a_{<}}{a_{>}}\right)^{\varrho}\; \mathcal{F}^0_{lmn\bar{p}}\; \mathcal{G}_{lmn\bar{p}\bar{q}}^A(e_A)\; \mathcal{G}_{lmn\bar{p}\bar{q}}^B(e_B)\; . \label{eq:clmnpq_B}
\end{equation}\\

In addition to the planet--planet tides, planet $A$ is also deformed by its tidal interaction with the host star, in which case we express the disturbing potential as

\begin{align}
\delta\mathcal{U}_{lm}^{*}(\mathbf{r}') &= \sum_{p=0}^{l} \sum_{q=-\infty}^{\infty} k_{lm}(\omega_{lmpq})\, \frac{R_{A}^{2l+1}}{(r')^{l+1}} \, B_{lm}^{*} \, C_{lmpq}^{*} \, \mathcal{P}_{lm}(\cos{\vartheta'})\;\times\nonumber\\[2ex]
&\hspace{3cm}\times\;
\cos
\Big[\nu_{lmpq}-\varepsilon_{lm}(\omega_{lmpq}) - m(\varphi'+\theta)\Big]
\label{eq:disturbingst}
\end{align}
with

\begin{equation}
\nu_{lmpq} = (l-2p+q)\,M_A + (l-2p)\,\varpi_A\; ,
\end{equation}\\
\begin{equation}
B_{lm}^{*} = \mathcal{G}\mathcal{M}_{*}\, \frac{(l-m)!}{(l+m)!}\, (2-\delta_{m0})\; ,
\end{equation}\\
and

\begin{equation}
C_{lmpq}^{*} = \frac{1}{a_{A}^{l+1}}\; \mathcal{F}_{lmp}(0)\; \mathcal{G}_{lpq}(e_A)\; .
\end{equation}\\
where $\mathcal{F}_{lmp}$ and $\mathcal{G}_{lpq}$ are Kaula's functions of inclination and eccentricity and $\mathcal{M}_{*}$ is the stellar mass.

In the model system consisting of three bodies---the host star and the two planets---we may consider four components to the tidal torque influencing planet $A$'s spin rate. The leading component is obviously the torque raised by the host star on the deformation induced by the star itself, which leads the planet either into a spin--orbit resonance or towards pseudo--synchronisation \citep[e.g.,][]{makarov13,ferrazmello13,correia14}. Second, there are two ``cross-components": planet $B$ acting on the tidal bulge raised by the host star, and vice versa. These two terms are presumably non--negligible only around the mutual conjunctions of the two planets. Last, there is also a tiny component due to planet $B$'s action on the deformation caused by the planet--planet tides.\\

The tidal torque raised on a planet with zero obliquity by the host star or by another body in the same orbital plane is orthogonal to the planetary equator. It induces changes in the spin rate (including physical libration) but has no effect on the axial tilt. Each term $\mathcal{T}_{lm}$ of the torque's harmonic expansion can be expressed as \citep{efroimsky09}

\begin{equation}
\mathcal{T}_{lm} = \mathcal{M}'\; \frac{\partial}{\partial\theta} \Big[\delta\mathcal{U}_{lm}(\mathbf{r}')\Big]\; , \label{eq:torquegen}
\end{equation}\\
where $\mathcal{M}'$ is mass of the spin--perturbing body located at position $\mathbf{r}'$.

On our way towards the derivation of all planetary contributions to the total tidal torque, we will begin with the cross-components. In order to express the disturbing potential sensed at the instantaneous position of planet $B$ due to the deformation of planet $A$ induced by stellar tides, we apply the harmonic expansion derived in the previous section on the terms of expression (\ref{eq:disturbingst}) containing $r'$, $\vartheta'$ and $\varphi'$. This results in

\begin{align}
\delta\mathcal{U}_{lm}^{*B}\equiv\delta\mathcal{U}_{lm}^{*}(\mathbf{r}')\bigg\rvert_{\mathbf{r'}=\mathbf{r}} &= \sum_{h=0}^{l} \sum_{j=-\infty}^{\infty} k_{lm}(\omega_{lmhj})\, R_{A}^{2l+1} \, B_{lm}^{*} \, C_{lmhj}^{*} \;\times \nonumber\\
&\hspace{0.5cm}\times\;\sum_{n}\sum_{\bar{p}} \sum_{\bar{q}} C_{lmn\bar{p}\bar{q}}^{B}\, \cos\Big[(\nu_{lmhj}-m\theta)-(\nu_{mn\bar{p}\bar{q}}-m\tilde{\theta})-\varepsilon_{lm}(\omega_{lmhj})\Big]\; , \label{eq:dustar}
\end{align}\\
where $\tilde{\theta}$ is the sidereal time with respect to the rotation of perturbed planet $A$. Note that $\theta$ and $\tilde{\theta}$ are essentially the same quantity and here \citep[as well as in][]{kaula64} they are labelled separately only for the sake of the following derivations. Now, we may differentiate (\ref{eq:dustar}) with respect to $\tilde{\theta}$, multiply the result by $m'=m_B$ and then set $\theta=\tilde{\theta}$. The general form of the tidal torque becomes

\begin{align}
\mathcal{T}_{lm}^{*B} &= -\mathcal{M}_B\; m\sum_{h=0}^{l} \sum_{j=-\infty}^{\infty} k_{lm}(\omega_{lmhj})\, R_{A}^{2l+1} \, B_{lm}^{*} \, C_{lmhj}^{*} \;\times\nonumber\\
&\hspace{4cm}\times\;\sum_{n}\sum_{\bar{p}} \sum_{\bar{q}} C_{lmn\bar{p}\bar{q}}^{B}\, \sin\Big[\nu_{lmhj}-\nu_{mn\bar{p}\bar{q}}-\varepsilon_{lm}(\omega_{lmhj})\Big]\; . \label{eq:tstar}
\end{align}\\
Equation (\ref{eq:tstar}) can be further split into a secular term, which affects the evolution of planet $A$'s spin rate over hundreds of orbital periods, and a periodic term, responsible for the physical libration. The secular term, which remains non--zero after averaging over the orbital periods of the two planets as well as over the period of apsidal precession, is characterised by the condition $\nu_{lmpq}=\nu_{mn\bar{p}\bar{q}}$. This leads to four sub--conditions:

\begin{equation}
\begin{gathered}
q_1 = j\;, \\[2ex]
q_2 = 0\;, \\[2ex]
p_1 = 2(p_2+p_4) - n\;, \\[2ex]
m - 2(p_2+p_3) = l - 2h\;.
\end{gathered} \label{eq:torque_cross_sec}
\end{equation}\\
Furthermore, in the coplanar case treated here, indices $l$, $m$, and $h$ are linked by the relation $m-l+2h = 0$ \citep{murray99}. Combined with the last line of equations (\ref{eq:torque_cross_sec}), this relation results in the additional condition:

\begin{equation}
p_2=-p_3\equiv0\; .
\end{equation}\\
Hence, the prefactor from equation (\ref{eq:clmnpq_B}) assumes a simplified form with the following substitutions:

\begin{equation}
\begin{gathered}
\mathcal{F}^{0, {\rm{sec}}}_{lmnp_4} = \frac{1}{2^{l+m}}\; \frac{(l+m)!}{l!}\; \binom{l}{\frac{l-m}{2}} (-1)^{\frac{l-m}{2}} \sum_{s=0}^{m} \binom{m}{s} \sum_{d_1} \binom{m-s}{2p_4-n-d_1} \binom{s}{d_1} (-1)^{2p_4-n} \; \times \\[2ex]
\times\; \sum_{u} (-1)^{u} \frac{\Gamma\left(n-u+\frac{l+m+1}{2}\right)}{\Gamma\left(\frac{l+m+1}{2}\right)\,u!\,(n-2u)!}\; \binom{n-2u}{p_4-u}\; ,
\end{gathered} \label{eq:fpref_sec}
\end{equation}\\
\begin{equation}
\begin{aligned}
    \mathcal{G}_{lmnp_4 j}^{A, {\rm{sec}}}(e_A) = X_{m+j}^{m+2(n-p_4),m}(e_A) \quad &\mbox{for inner planet A} \\[2ex]
    \mathcal{G}_{lmnp_4 j}^{A, {\rm{sec}}}(e_A) = X_{m+j}^{-2p_4-l-1,m}(e_A) \quad &\mbox{for outer planet A}
\end{aligned}
\end{equation}\\
and

\begin{equation}
\begin{aligned}
    \mathcal{G}_{lmnp_4 j}^{B, {\rm{sec}}}(e_B) = X_{0}^{-m-2(n-p_4)-l-1,0}(e_B) \quad &\mbox{for inner planet A} \\[2ex]
    \mathcal{G}_{lmnp_4 j}^{B, {\rm{sec}}}(e_B) = X_{0}^{2p_4,0}(e_B) \quad &\mbox{for outer planet A}
\end{aligned}
\end{equation}\\
In equation (\ref{eq:fpref_sec}), only the terms with $2p_4-n\leq m$ are non-zero. Similar expressions can also be derived for the other cross-component, arising due to the action of the host star on the tidal bulge raised on planet $A$ by planet $B$. Although the variations of the two cross-components over the timescale of an orbital period differ, their secular part is the same: $\mathcal{T}_{lm}^{*B, {\rm{sec}}}=\mathcal{T}_{lm}^{B*, {\rm{sec}}}$.\newline

The remaining, planet--planet component can be derived from the disturbing potential

\begin{align}
\delta\mathcal{U}_{lm}^{BB}\equiv\delta\mathcal{U}_{lm}^{B}(\mathbf{r}')\bigg\rvert_{\mathbf{r'}=\mathbf{r}} &= \sum_{n} \sum_{\bar{p}} \sum_{\bar{q}} k_{lm}(\omega_{mn\bar{p}\bar{q}})\, R_{A}^{2l+1} \, B_{lm}^{B} \, C_{lmn\bar{p}\bar{q}}^{B} \; \times \nonumber\\[2ex]
\hspace{-1cm}&\times\;\sum_{N}\sum_{\bar{P}} \sum_{\bar{Q}} C_{lmN\bar{P}\bar{Q}}^{B}\, \cos\Big[(\nu_{mn\bar{p}\bar{q}}-m\theta)-(\nu_{mN\bar{P}\bar{Q}}-m\tilde{\theta})-\varepsilon_{lm}(\omega_{mn\bar{p}\bar{q}})\Big]\; .
\label{eq:duplpl}\end{align}
\\
in the same manner. Here, the multiindices $\bar{P}$ and $\bar{Q}$ are defined similarly to $\bar{p}$ and $\bar{q}$ as $\bar{P}=\{P_1,P_2,P_3,P_4\}$ and $\bar{Q}=\{Q_1,Q_2\}$. We obtain the planet-planet tidal torque

\begin{align}
\mathcal{T}_{lm}^{BB} &= -m_B\; m\sum_{n} \sum_{\bar{p}} \sum_{\bar{q}} k_l(\omega_{lmn\bar{p}\bar{q}})\, R_{A}^{2l+1} \, B_{lm}^{B} \, C_{lmn\bar{p}\bar{q}}^{B} \;\times\nonumber\\ &\hspace{3cm}\times\;\sum_{N}\sum_{\bar{P}} \sum_{\bar{Q}} C_{lmN\bar{P}\bar{Q}}^{B}\, \sin\Big[\nu_{lmn\bar{p}\bar{q}}-\nu_{lmN\bar{P}\bar{Q}}-\varepsilon_{lmn\bar{p}\bar{q}}\Big]\; , \label{eq:tplpl}
\end{align}\\
whose secular terms satisfy the following conditions

\begin{equation}
\begin{gathered}
Q_1 = q_1\;, \\[2ex]
Q_2 = q_2\;, \\[2ex]
P_2+P_3 = p_2+p_3\;,\\[2ex]
N+P_1-2(P_2+P_4) = n+p_1-2(p_2+p_4)\;.
\end{gathered} \label{eq:tplpl_constr}
\end{equation}\\
In this case, the secular term has a more complex form than in the case of cross-components.

\section{Consequences for the orbital evolution} \label{sec:evol}

Rotational variations caused by the tidal interaction, derived in the previous section, go hand in hand with variations of orbital elements. Since we have limited the discussion to a coplanar case, our set of orbital elements consists of the semi-major axes $a_{\rm{A}}$, $a_{\rm{B}}$, the eccentricities $e_{\rm{A}}$, $e_{\rm{B}}$, and the longitudes of pericenter $\varpi_{\rm{A}}$, $\varpi_{\rm{A}}$. Their variations under the action of a disturbing function $\mathcal{R}$, equal in this case to the additional potential arising from the surface deflections of the tidally loaded bodies, equations (\ref{eq:dustar}) and (\ref{eq:duplpl}), is described by Lagrange planetary equations \citep[e.g.,][]{murray99}:

\begin{align}
    \ddt{a_k} &= \frac{2}{n_k a_k} \prp{M_k}\; , \label{eq:lagrange1} \\[0.5em]
    \ddt{e_k} &= \frac{1-e_k^2}{n_k a_k^2 e_k} \prp{M_k} - \frac{\sqrt{1-e_k^2}}{n_k a_k^2 e_k} \prp{\varpi_k}\; , \label{eq:lagrange2} \\[0.5em]
    \ddt{\varpi_k} &= \frac{\sqrt{1-e_k^2}}{n_k a_k^2 e_k}\; , \prp{e_k} \label{eq:lagrange3}
\end{align}\\
where the subscript $k$ stands for $A$ or $B$. Their result is a correction that can be added to the evolution of orbital elements caused by mutual gravitational (third-body) perturbations and by the planetary oblateness (or any other density anomalies). Here, we will illustrate the form of the partial derivatives in equations (\ref{eq:lagrange1}-\ref{eq:lagrange3}) for the case of planet $A$ perturbed by planet $B$ and the primary. 

As with the tidal torque, the planet-planet tidal contribution to the evolution equations can be divided into three components. First, the deformation of planet $A$ caused by planet $B$ affects the evolution of planet $A$'s orbit around the primary. Second, the deformation of planet $A$ caused by the primary contributes to the evolution of planet $B$'s orbit around the primary. Finally, the deformation of planet $A$ caused by planet $B$ also affects the orbit of planet $B$ (this is the purely planet-planet component).\\

For the first cross-term, we can write the following partial derivatives:

\begin{align}
    \frac{\partial}{\partial M_{A}} \Big[\delta\mathcal{U}^{B*}_{lm}\Big] &= \sum_{n} \sum_{\bar{h}} \sum_{\bar{j}} k_{lm}(\omega_{mn\bar{h}\bar{j}})\, R_{A}^{2l+1} \, B_{lm}^{B} \, C_{lmn\bar{h}\bar{j}}^{B} \;\times \nonumber\\
    &\times\;\sum_{p} \sum_{q} (l-2p+q) C_{lmpq}^{*}\, \sin\Big[\nu_{mn\bar{h}\bar{j}}-\nu_{lmpq}-\varepsilon_{lm}(\omega_{mn\bar{h}\bar{j}})\Big]\; , \label{eq:prpm_ba}
\end{align}\\
\begin{align}
    \frac{\partial}{\partial \varpi_{A}} \Big[\delta\mathcal{U}^{B*}_{lm}\Big] &= \sum_{n} \sum_{\bar{h}} \sum_{\bar{j}} k_{lm}(\omega_{mn\bar{h}\bar{j}})\, R_{A}^{2l+1} \, B_{lm}^{B} \, C_{lmn\bar{h}\bar{j}}^{B} \;\times \nonumber\\
    &\times\;\sum_{p} \sum_{q} (l-2p) C_{lmpq}^{*}\, \sin\Big[\nu_{mn\bar{h}\bar{j}}-\nu_{lmpq}-\varepsilon_{lm}(\omega_{mn\bar{h}\bar{j}})\Big]\; , \label{eq:prpp_ba}
\end{align}\\
and
\begin{align}
    \frac{\partial}{\partial e_{A}} \Big[\delta\mathcal{U}^{B*}_{lm}\Big] &= \sum_{n} \sum_{\bar{h}} \sum_{\bar{j}} k_{lm}(\omega_{mn\bar{h}\bar{j}})\, R_{A}^{2l+1} \, B_{lm}^{B} \;\times \nonumber\\
    &\times\;\sum_{p} \sum_{q} \left[\frac{\partial C_{lmn\bar{h}\bar{j}}^{B}}{\partial e_A} C_{lmpq}^{*} + C_{lmn\bar{h}\bar{j}}^{B} \frac{\partial C_{lmpq}^{*}}{\partial e_A}\right]\, \cos\Big[\nu_{mn\bar{h}\bar{j}}-\nu_{lmpq}-\varepsilon_{lm}(\omega_{mn\bar{h}\bar{j}})\Big]\; . \label{eq:prpe_ba}
\end{align}\\

In the second cross-component, we are considering the effect of a deformed planet $A$ on the orbit of planet $B$. Therefore, we have to differentiate the disturbing function with respect to $M_{B}$, $\varpi_B$, and $e_B$:

\begin{align}
    \frac{\partial}{\partial M_{B}} \Big[\delta\mathcal{U}^{*B}_{lm}\Big] &= \sum_{h} \sum_{j} k_{lm}(\omega_{lmhj})\, R_{A}^{2l+1} \, B_{lm}^{*} \, C_{lmhj}^{*} \;\times \nonumber\\
    &\times\;\sum_{n} \sum_{\bar{p}} \sum_{\bar{q}} (p_1-2p_2+n-2p_4+q_2) C_{lmn\bar{p}\bar{q}}^{B}\, \sin\Big[\nu_{lmhj}-\nu_{mn\bar{p}\bar{q}}-\varepsilon_{lm}(\omega_{lmhj})\Big]\; , \label{eq:prpm_ab}
\end{align}\\
\begin{align}
    \frac{\partial}{\partial \varpi_{B}} \Big[\delta\mathcal{U}^{*B}_{lm}\Big] &= \sum_{h} \sum_{j} k_{lm}(\omega_{lmhj})\, R_{A}^{2l+1} \, B_{lm}^{*} \, C_{lmhj}^{*} \;\times \nonumber\\
    &\times\;\sum_{n} \sum_{\bar{p}} \sum_{\bar{q}} (p_1-2p_2+n-2p_4) C_{lmn\bar{p}\bar{q}}^{B}\, \sin\Big[\nu_{lmhj}-\nu_{mn\bar{p}\bar{q}}-\varepsilon_{lm}(\omega_{lmhj})\Big]\; , \label{eq:prpp_ab}
\end{align}\\
and
\begin{align}
    \frac{\partial}{\partial e_{B}} \Big[\delta\mathcal{U}^{*B}_{lm}\Big] &= \sum_{h} \sum_{j} k_{lm}(\omega_{lmhj})\, R_{A}^{2l+1} \, B_{lm}^{*} \, C_{lmhj}^{*} \;\times \nonumber\\
    &\times\;\sum_{n} \sum_{\bar{p}} \sum_{\bar{q}} \frac{\partial C_{lmn\bar{p}\bar{q}}^{B}}{\partial e_B}\, \cos\Big[\nu_{lmhj}-\nu_{mn\bar{p}\bar{q}}-\varepsilon_{lm}(\omega_{lmhj})\Big]\; . \label{eq:prpe_ab}
\end{align}\\
The partial derivatives in the (most complicated) planet-planet component are also taken with respect to $M_B$, $\varpi_B$, and $e_B$. They read as:

\begin{align}
    \frac{\partial}{\partial M_{B}} \Big[\delta\mathcal{U}^{BB}_{lm}\Big] &= \sum_{n} \sum_{\bar{p}} \sum_{\bar{q}} k_{lm}(\omega_{mn\bar{p}\bar{q}})\, R_{A}^{2l+1} \, B_{lm}^{B} \, C_{lmn\bar{p}\bar{q}}^{B} \;\times \nonumber\\
    &\times\;\sum_{w} \sum_{\bar{h}} \sum_{\bar{j}} (h_1-2h_2+w-2h_4+j_2) C_{lmw\bar{h}\bar{j}}^{B}\, \sin\Big[\nu_{mn\bar{p}\bar{q}}-\nu_{mw\bar{h}\bar{j}}-\varepsilon_{lm}(\omega_{mn\bar{p}\bar{q}})\Big]\; , \label{eq:prpm_bb}
\end{align}\\
\begin{align}
    \frac{\partial}{\partial \varpi_{B}} \Big[\delta\mathcal{U}^{BB}_{lm}\Big] &= \sum_{n} \sum_{\bar{p}} \sum_{\bar{q}} k_{lm}(\omega_{mn\bar{p}\bar{q}})\, R_{A}^{2l+1} \, B_{lm}^{B} \, C_{lmn\bar{p}\bar{q}}^{B} \;\times \nonumber\\
    &\times\;\sum_{w} \sum_{\bar{h}} \sum_{\bar{j}} (h_1-2h_2+w-2h_4) C_{lmw\bar{h}\bar{j}}^{B}\, \sin\Big[\nu_{mn\bar{p}\bar{q}}-\nu_{mw\bar{h}\bar{j}}-\varepsilon_{lm}(\omega_{mn\bar{p}\bar{q}})\Big]\; , \label{eq:prpp_bb}
\end{align}\\
and

\begin{align}
    \frac{\partial}{\partial e_{B}} \Big[\delta\mathcal{U}^{BB}_{lm}\Big] &= \sum_{n} \sum_{\bar{p}} \sum_{\bar{q}} k_{lm}(\omega_{mn\bar{p}\bar{q}})\, R_{A}^{2l+1} \, B_{lm}^{B} \, \;\times \nonumber\\
    &\times\;\sum_{w} \sum_{\bar{h}} \sum_{\bar{j}} \left[\frac{\partial C_{lmn\bar{p}\bar{q}}^{B}}{\partial e_B} C_{lmw\bar{h}\bar{j}}^{B} + C_{lmn\bar{p}\bar{q}}^{B}\frac{\partial C_{lmw\bar{h}\bar{j}}^{B}}{\partial e_B}\right]\, \cos\Big[\nu_{mn\bar{p}\bar{q}}-\nu_{mw\bar{h}\bar{j}}-\varepsilon_{lm}(\omega_{mn\bar{p}\bar{q}})\Big]\; . \label{eq:prpe_bb}
\end{align}\\
Note that all tidal interaction is mutual. Although we are only listing the components of the evolution equations and the tidal torque arising from the deformation of planet $A$, in practical applications, we should also include the effect of planet $B$'s deformation on the evolution of both planetary orbits and on planet $B$'s spin rate.

\section{Comparison with a numerical solution} \label{sec:comparison}

To ensure that the analytical expressions for planet-planet tidal potential, tidal torque, and variations of Keplerian elements produce a behaviour replicable by an independent method, we implemented the mutual tidal loading to a numerical tool similar to the one used in \citet{walterova17}. The numerical tool solves the governing equations of a viscoelastic spherical shell subjected to a volumetric force, applying spherical harmonic decomposition in the lateral directions and staggered-grid finite difference method in the radial direction. Instantaneous positions of the two planets with respect to the host star are obtained by standard methods of celestial mechanics \citep[e.g.,][]{murray99}. Details of the implementation are given in Appendix \ref{sec:appA}.

\subsection{Tidal potential}

First, we compare the tidal potential on planet $A$'s surface obtained with equation (\ref{eq:potecc}) with a numerically calculated value,

\begin{equation}
    \mathcal{U}(R_{A},\vartheta',\varphi') = \frac{\mathcal{G} \mathcal{M}_B}{r} \sum_{l=2}^{+\infty} \left(\frac{R_{A}}{r}\right)^l \sum_{m=0}^l (2-\delta_{m0}) \frac{(l-m)!}{(l+m)!} \mathcal{P}_{lm}(\cos{\vartheta'}) \mathcal{P}_{lm}(\cos{\vartheta}) \cos{[m(\varphi'-\varphi)]}\; .
\end{equation}\\
The instantaneous distance of planet $B$ from planet $A$, denoted as $r$, and the planetographic longitude of the sub-$B$ point on $A$'s surface, denoted as $\varphi$, are calculated according to Appendix \ref{sec:appA}. Colatitude $\vartheta$ is, in a coplanar case, identically zero. The potential is evaluated at coordinates $(r',\vartheta',\varphi')=(R_{A},\frac{\pi}{2},0)$.

To evaluate equation (\ref{eq:potecc}) and, consequently, the full tidal potential,

\begin{equation}
    \mathcal{U} = \sum_{l=2}^{+\infty} \sum_{m=0}^{l} \mathcal{U}_{lm}\; ,
\end{equation}\\
we have to choose the maximum degrees $l_{\rm{max}}$ and $n_{\rm{max}}$ as well as the maximum eccentricity indices $q_{1,\rm{max}}$ and $q_{2,\rm{max}}$. Due to relatively small orbital eccentricities and large distances between the considered bodies, compared to $R_{A}$, the upper limits are set to: $l_{\rm{max}}=2$, $q_{1,\rm{max}}=q_{2,\rm{max}}=2$. For $n_{\rm{max}}$, we test several values between $10$ and $40$. The required degree strongly depends on the considered system, in particular on the maximum possible ratio of the shorter radial distance to the longer one,

\begin{equation}
    \alpha' = \max{\frac{r_{<}}{r_>}} = \frac{a_<(1+e_<)}{a_>(1-e_>)}\; ,
\end{equation}\\
where $a_<$ is the semi-major axis of the inner orbit, $e_<$ the eccentricity of the inner orbit, and $a_>$, $e_>$ are the same quantities for the outer orbit.

\begin{figure}
    \centering
    \includegraphics[width=1\textwidth]{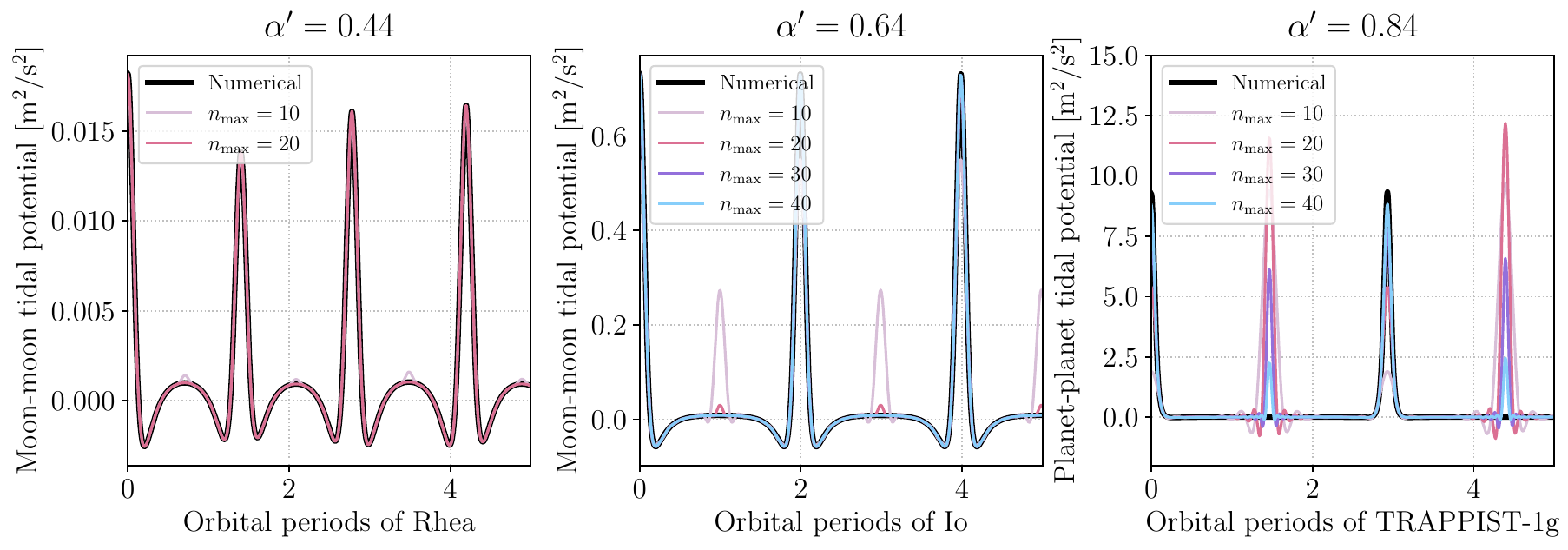}
    \caption{Three examples of planet-planet (or moon-moon) tidal potential sensed at coordinates $(r',\vartheta',\varphi')=(R_{\rm{surf}},\frac{\pi}{2},0)$ of ``planet $A$". Thick black line indicates the potential obtained with the numerical code. Different shades of the thinner lines correspond to different choices of the upper limit on index $n$ in the implementation of the analytical formula. \textit{Left:} Titan acting on Rhea, \textit{Middle:} Europa acting on Io, \textit{Right:} TRAPPIST-1f acting on TRAPPIST-1g.}
    \label{fig:potentials}
\end{figure}

Figure \ref{fig:potentials} illustrates the time variations of the tidal potential on the surface of ``planet $A$" over its five orbital periods. The role of planet $A$ is played by the moons Rhea (left panel), Io (middle panel) or the exoplanet TRAPPIST-1g (right panel), and planet $B$ is embodied by Titan, Europa or TRAPPIST-1f, respectively. Parameters of the model systems are summarised in Table \ref{tab:model_params}. In all cases, the observed variations show peaks at each conjunction between the two moons or planets.

\begin{table}[]
    \centering
    \begin{tabular}{| l | l | l | l |}
        \hline
        Parameter & Rhea \& Titan & Io \& Europa & TRAPPIST-1 g \& f \\
        \hline
        $m_*$ & $\unit[5.68\times10^{26}]{kg}$ & $\unit[1.90\times10^{27}]{kg}$ & $0.09\; m_{\rm{Sun}}$ \\[0.3em]
        $m_A$ & $\unit[2.31\times10^{21}]{kg}$ & $\unit[8.93\times10^{22}]{kg}$ & $1.15\; m_{\rm{Earth}}$ \\[0.3em]
        $m_B$ & $\unit[1.35\times10^{23}]{kg}$ & $\unit[4.80\times10^{22}]{kg}$ & $0.93\; m_{\rm{Earth}}$ \\[0.3em]
        $a_A$ & $\unit[5.27\times10^{8}]{m}$ & $\unit[4.22\times10^{8}]{m}$ & $\unit[0.047]{AU}$ \\[0.3em]
        $a_B$ & $\unit[12.22\times10^{8}]{m}$ & $\unit[6.71\times10^{8}]{m}$ & $\unit[0.039]{AU}$ \\[0.3em]
        $e_A$ & $0.001$ & $0.004$ & $0.0021$ \\[0.3em]
        $e_B$ & $0.029$ & $0.009$ & $0.0101$ \\[0.3em]
        $R_A$ & $\unit[763.5]{km}$ & $\unit[1821.5]{km}$ & $\unit[1.148]{R_{\rm{Earth}}}$\\[0.3em]
        \hline
    \end{tabular}
    \caption{Overview of the parameters used for the comparison of analytical and numerical approach. The longitudes of periastron were, in all cases, set to zero ($\varpi_A=\varpi_B=0$). Parameters of the Saturnian satellites were adopted from \citet{jacobson2022}, the Jovian satellites from \citet{jacobson2021}, and the TRAPPIST-1 planets from \citet{grimm2018}.}
    \label{tab:model_params}
\end{table}

For Rhea and the outer perturber Titan, we have $\alpha'=0.44$. Summation of equation (\ref{eq:potecc}) up to $n_{\rm{max}}=10$ leads to a relatively good correspondence with the numerically calculated potential except for small artificial peaks in the middle of the conjunction periods. The assumption of $n_{\rm{max}}=20$ yields a perfect alignment of the numerical and the analytical solution. For Io disturbed by the outer moon Europa, with $\alpha'=0.64$, the expansion already requires $n_{\rm{max}}=30$ or, optimally, $n_{\rm{max}}=40$. Finally, for TRAPPIST-1g disturbed by the inner planet TRAPPIST-1f, we have $\alpha'=0.84$ and the expansion requires $n_{\rm{max}}>40$. Interestingly, a low $n_{\rm{max}}$ in this case predicts false maxima in between the conjunctions while strongly underestimating the true maxima.

Our naïve implementation of the analytical formulae is numerically unstable at $n>40$ and, for this reason, we were unable to fully reproduce the true time dependence of the planet-planet tidal potential sensed by TRAPPIST-1g. This issue is caused by the summation of ratios of gamma functions in equation (\ref{eq:fpref}) and its elimination would require a dedicated effort, outside the scope of this paper.

\subsection{Tidal torque}

For the sake of simplicity, we illustrate the individual components of the tidal torque on the specific case of the moon Rhea, tidally interacting with Saturn and the outer moon Titan. As we have seen in the previous section, the expansion of tidal potential can be truncated at $n_{\rm{max}}=20$. The result is compared against the numerical solution from Appendix \ref{sec:appA}.

To evaluate the tidal torque, it is necessary to adopt a rheological model of the planet as well as corresponding rheological parameters. We will assume that the planet is governed by the viscoelastic Maxwell rheology and that the complex tidal Love number $\bar{k}_{l}(\chi)=k_l(\chi) e^{-\iu\varepsilon_l(\chi)}$ is \citep[e.g.,][]{castillo11,efroimsky12}

\begin{equation}
    \bar{k}_{l}(\chi) = \frac{3}{2(l-1)} \frac{1}{1+\frac{2l^2+4l+3}{l}\frac{\bar{\mu}(\chi)}{\rho g R_A}}\; ,
\end{equation}\\
where, in the case of tidal loading by planet $B$, $\chi=\chi_{mn\bar{p}\bar{q}}=|\omega_{mn\bar{p}\bar{q}}|$ is the (unsigned) physical frequency corresponding to mode $\{mn\bar{p}\bar{q}\}$, $\rho$ represents the average density of planet $A$, and $g$ is its surface gravity. For tidal loading by the primary, we assume $\chi=\chi_{lmpq}=|\omega_{lmpq}|$. Symbol $\bar{\mu}(\chi)$ stands for the complex rigidity and contains information about the rheological model. In the case of Maxwell rheology, the inverse value of complex rigidity attains the following form:

\begin{equation}
    \frac{1}{\bar{\mu}(\chi)} = \frac{1}{\mu} - \frac{\iu}{\eta\chi}\; .
\end{equation}\\
Here, $\mu$ represents the unrelaxed rigidity and $\eta$ is the viscosity.\\

\begin{figure}
    \centering
    \includegraphics[width=1\textwidth]{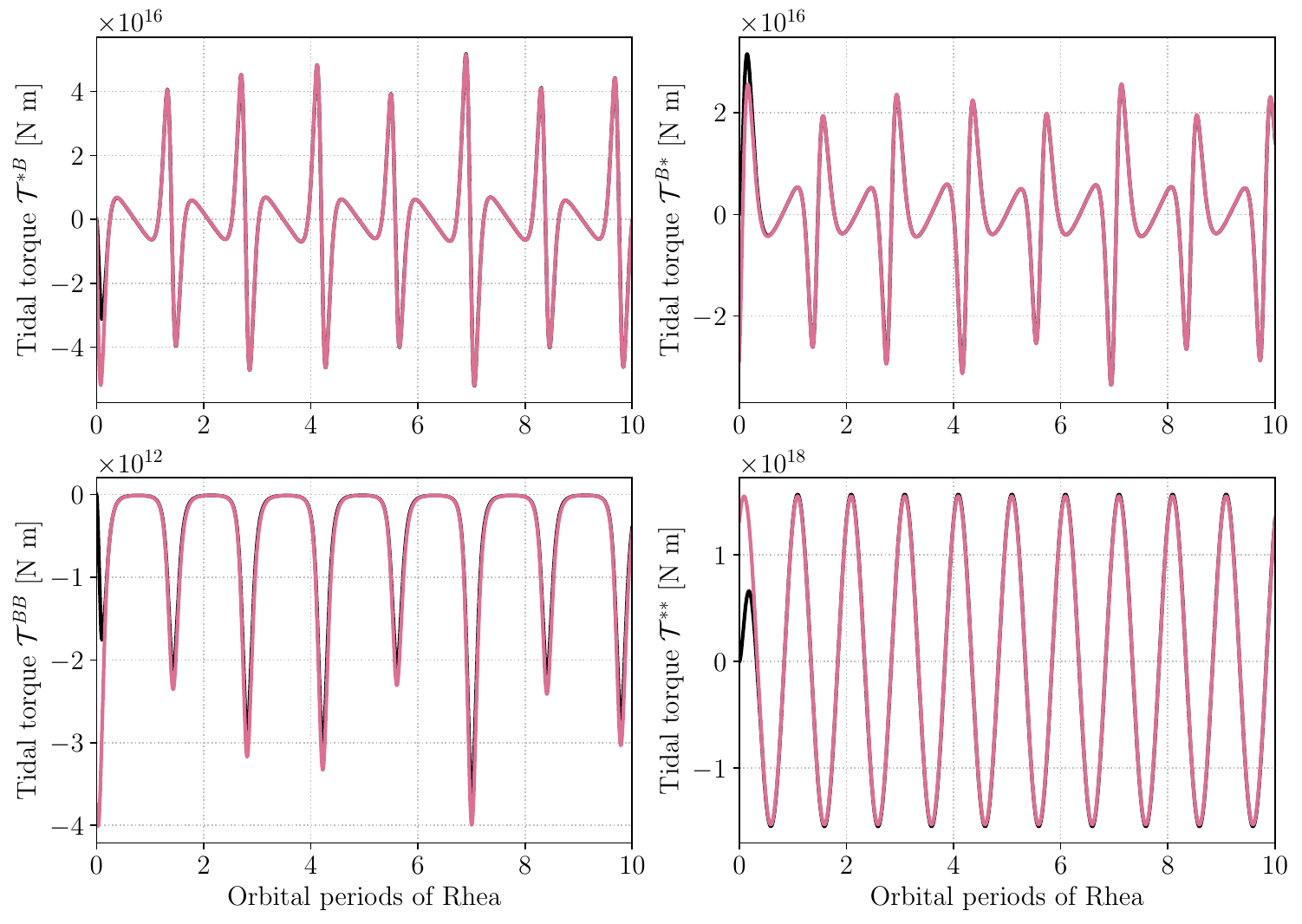}
    \caption{Illustration of the individual components of the tidal torque acting on Rhea due to tidal loading by Saturn and Titan. Black line marks the numerical solution, pink line is the analytical solution calculated according to Section \ref{sec:pl-pl-torque}, with $n_{\rm{max}}=20$.}
    \label{fig:torques}
\end{figure}

Before discussing the comparison of the analytical and the numerical solution, we should emphasize an inevitable difference between the two approaches: the numerical solution naturally includes transient phenomena, in particular the viscoelastic relaxation of the studied moon or planet. As an example, when subjected to a time-independent volumetric force (e.g., the constant part of the tidal force), the viscoelastic body in the numerical model will gradually relax to the hydrostatic shape, with a rate of this adjustment governed by $\eta$. In the analytical model, on the other hand, the studied body will immediately act as having the hydrostatic shape.

To facilitate the comparison and reduce the time needed to reach a relaxed shape in the numerical model, we prescribe a very low effective viscosity of Rhea, $\eta=\unit[10^{12}]{Pa\,s}$. The rigidity is set to $\mu=\unit[5]{GPa}$.\\

The effect of initial viscoelastic relaxation is noticeable at the beginning of the time series for the individual components of the tidal torque depicted in Figure \ref{fig:torques}. For the rest of the evolution, the analytical and the numerical solution present almost identical behaviour. Small differences can be seen in the peak values of the planet-planet component $\mathcal{T}^{BB}$ and of the primary-induced component $\mathcal{T}^{**}$. These deviations are likely related to the differences in the two implementations.

In Figure \ref{fig:torques}, we can also compare the magnitudes of the individual tidal torque components. The peak values of the cross-components are, in this particular case, two orders of magnitude lower than the peak values of the primary-induced component $\mathcal{T}^{**}$. The planet-planet torque $\mathcal{T}^{BB}$ is much smaller, and its peak values are six orders of magnitude lower than $\mathcal{T}^{**}$. The mean value of all components related to planet-planet tides ($\mathcal{T}^{*B}$, $\mathcal{T}^{B*}$, and $\mathcal{T}^{BB}$) is $\sim\unit[10^{11}]{N\,m}$ and it is four orders of magnitude smaller than the mean value of $\mathcal{T}^{**}$.

Note that in these tests, we are only considering tidal deformation, and not the permanent, possibly non-hydrostatic shape of Rhea \citep{mackenzie2008}. The torque acting on the permanent shape of the moon both from the other moon and the primary would likely be greater than the components considered here \citep[e.g.,][]{rambaux2010} and should be included in practical applications.

\subsection{Semi-major axis}

\begin{figure}
    \centering
    \includegraphics[width=1\textwidth]{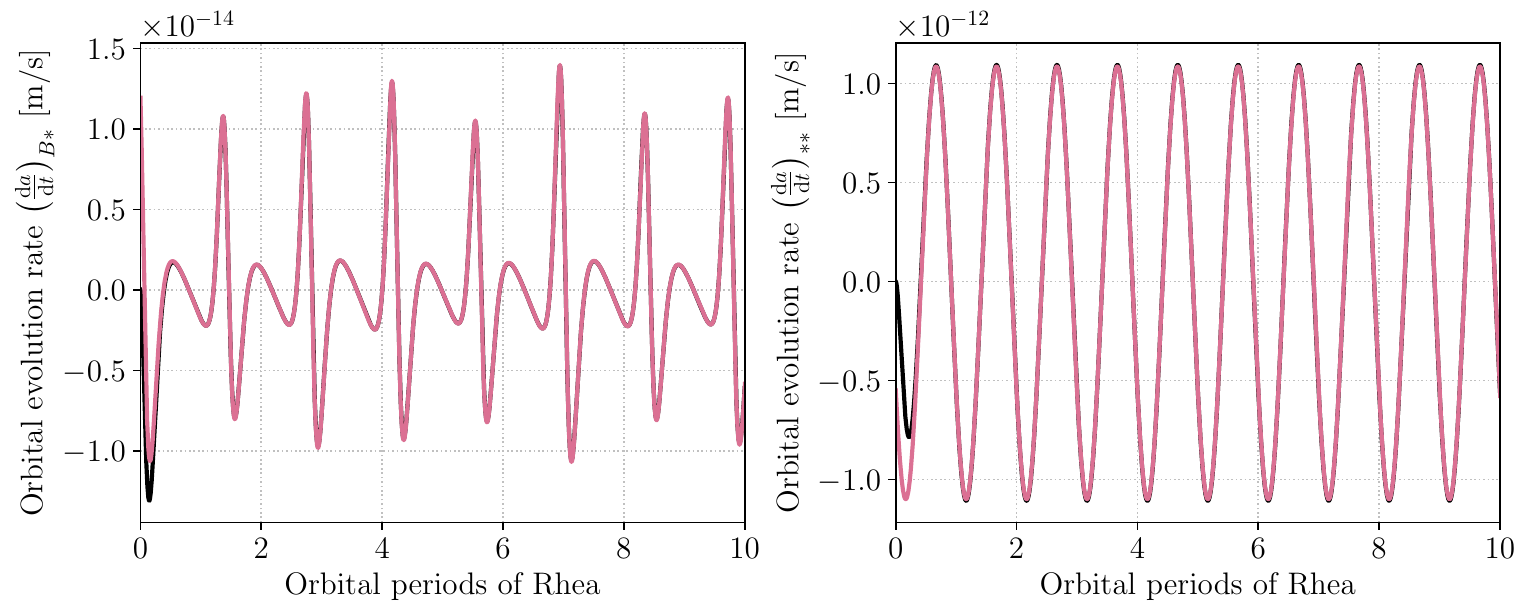}
    \caption{Illustration of Rhea's semi-major axis evolution. \textit{Left:} the effect of Titan-induced deformation. \textit{Right:} the effect of Saturn-induced deformation. As in Figure \ref{fig:torques}, the numerical solution is plotted in black and the analytical solution in pink.}
    \label{fig:semimajors}
\end{figure}

In the last comparison, we focus on the semi-major axis of Rhea that varies due to the moon's deformation caused by Titan and Saturn. The rheological model and $n_{\rm{max}}$ are the same as in the previous case. Figure \ref{fig:semimajors} shows a similar trend as Figure \ref{fig:torques}, including the small discrepancy between the peak values of the primary-induced evolution calculated with the analytical and the numerical code, respectively. Increasing spin rate (positive $\mathcal{T}$) is typically associated with a decreasing semi-major axis (negative ${\rm{d}}a/{\rm{d}}t$) and vice versa. The peak values of the $B*$ component are, again, two orders of magnitude lower than for the $**$ component.

Calculation of the other orbital evolution rates follows a similar procedure and is not illustrated here.

\section{Discussion and Conclusions} \label{sec:conclusion}

We have introduced a series expansion for the tidal potential arising from the mutual interaction between two planets or moons orbiting the same primary in a coplanar configuration. Building on previous works \citep{kaula61,kaula64,efroimsky09}, we then derived the expressions for tidal torque and for the evolution of Keplerian elements.

The individual terms of the planet-planet tidal potential contain nine indices, in contrast to the six indices of the tidal potential raised by the primary \citep[][considered is the general case before averaging]{kaula61,kaula64}. This complexity might not always be desirable in practical applications, especially for mutually close orbits with a large ratio of the semi-major axes, $a_{<}/a_{>}$, where index $n$ reaches very high values ($>50$). The presence of $n$ in the argument of gamma functions and factorials in prefactor $\mathcal{F}^0_{lmn\bar{p}}$, equation (\ref{eq:fpref}), can easily lead to numerical instabilities, as illustrated in Figure \ref{fig:potentials}. Moreover, the calculation of the tidal torque and the orbital evolution rates from our analytical formulae may be more time-demanding than the numerical model from Appendix \ref{sec:appA}. Although the prefactors for each tidal frequency can be precalculated and many terms are equal to zero, we are still left with the multiplication of long series (especially for the purely planet-planet component and large $n$). In our implementation, the analytical calculation can be more than a factor of ten longer than the numerical alternative. A numerical solution, such as the one presented by \citet{hay19} or in our Appendix \ref{sec:appA}, is then preferable for large $a_{<}/a_{>}$.

On the other hand, the analytical approach enables direct inspection of the tidal frequencies involved in planet-planet loading as well as a relatively simple extraction of the secular term, which is faster than in a numerical model. Additionally, since the analytical solution uses a rheological law formulated in the frequency domain, it does not need to account for the initial relaxation period.\footnote{However, if the inclusion of transient phenomena is needed, the numerical solution should be used.}

As was briefly discussed in the Introduction, the planet-planet tides may be particularly important in compact exoplanetary systems \citep{lingam18,wright18b,hay19}, especially if the planets are synchronously rotating and their orbits are close to circular. Under these circumstances, the tides raised by the primary remain almost constant, while mutual tidal interaction between secondaries introduces modulation and energy dissipation \citep{wright18b}. In the solar system, moon-moon tides might help to probe the interior structures of the large moons of Jovian---and potentially also Saturnian---satellites \citep{vanhoolst2024}, provided that the moon's reaction is resonantly amplified by its subsurface ocean \citep{hay20,hay2022}. Another example in which our development might be helpful is the case of tides raised by a star on the moon of its planetary companion. This phenomenon may further contribute to the complex dynamics of such configurations \citep{makarov2023,makarov2025}. Lastly, a special case of ``planet-planet" tides, requiring proper treatment of the orbital variations, is the tidal evolution of a planet (or a moon) in a circumbinary system \citep[e.g.,][]{correia2016,zoppetti2021}.

The specific applications of our formulae are, however, left for future work.


\begin{acknowledgments}
This work was initially inspired by a discussion of one of the authors (M.W.) with Emeline Bolmont at EPSC 2018. Its preliminary version was included in M.W.'s doctoral thesis, defended in 2021. The development of the extended version was supported by Czech Science foundation project No. 26-21877S. We are sincerely grateful to Michael Efroimsky and Hamish Hay for their encouragement on our long journey to completing and submitting the study. M.W. would also like to express her gratitute to Ondřej Kureš and Alexandr Dizov for their efforts to improve the implementation of the analytical formulae and for willingness to discuss new ideas.
\end{acknowledgments}

\begin{contribution}

MW came up with the initial research concept, developed the analytical model, wrote the software and the manuscript, and plotted and analysed the results. MB contributed to the software, edited the manuscript, obtained the funding, and was responsible for supervision. Both authors discussed the implications of the work and were involved in the validation of results.


\end{contribution}

%



\appendix

\section{Description of the numerical model} \label{sec:appA}

Deformation of the viscoelastic spherical shell is calculated in the time domain by solving the following set of governing equations:

\begin{equation}
\nabla\cdot\mathbf{u}\,=\,0\, , \label{eq:cont}
\end{equation}

\begin{equation}
-\nabla\pi + \nabla\cdot\mathbf{D}=\,-\rho\,(\mathbf{f}_{\rm{t}}+\mathbf{f}_{\rm{sg}})\, , \label{eq:mom}
\end{equation}

\begin{equation}
2\mathbf{\varepsilon} = \frac{1}{\mu}\mathbf{D} + \int_0^t \frac{1}{\eta}\mathbf{D}(t')\,{\rm{d}}t' \label{eq:reo1}
\end{equation}\\
where $\mathbf{u}$ denotes the displacement vector, $\mathbf{\varepsilon}=\frac12(\nabla\mathbf{u}+\nabla^T\mathbf{u})$ is the incremental strain tensor, $\pi$ and $\mathbf{D}$ are the isotropic and the deviatoric parts of the incremental Cauchy stress tensor, $\mathbf{f}_{\rm{t}}$ and $\mathbf{f}_{\rm{sg}}$ represent the tidal and self-gravity force per unit mass, and $\rho$, $\eta$, and $\mu$ are the shell's average density, viscosity and rigidity (shear modulus), respectively. The first two equations embody the conservation of mass and linear momentum; the third one is the constitutive equation of Maxwell viscoelastic solid.

At the upper boundary, we assume free surface, while at the lower boundary, we prescribe the equilibrium of forces. We perform spherical harmonic decomposition of $\mathbf{u}$, $\mathbf{D}$, $\pi$, $\mathbf{f}_{\rm{t}}$, and $\mathbf{f}_{\rm{sg}}$. Radially-dependent equations (\ref{eq:cont}-\ref{eq:reo1}) for the individual spherical harmonic coefficients are then solved by finite difference method with a staggered grid. For details of the implementation, see \citet{walterova17}. On the following lines, we will focus on the inclusion of the planet-planet tides.

It should be noted that the code calculates the response of a spherical shell, or a viscoelastic mantle encompassing a liquid core. On the other hand, the tests presented in Section \ref{sec:comparison} assume a homogeneous planet. To obtain the closest possible correspondence between the two approaches, we prescribe here a small core ($R_{\rm{core}}=\unit[1]{m}$) and an identical density of the core and the mantle.\\

The tidal force sensed at a point $(r', \vartheta', \varphi')$ within the shell due to a perturber of mass $\mathcal{M}_{\rm{pert}}$ located at coordinates $(r_{\rm{pert}}, \vartheta_{\rm{pert}}, \varphi_{\rm{pert}})$ from the studied planet's center can be expanded into spherical harmonics as:

\begin{equation}
    \mathbf{f}_{\rm{t}} (r', \vartheta', \varphi') = \sum_{lm} f_{lm}^{l-1}(r')\, \mathbf{Y}_{lm}^{l-1} (\vartheta', \varphi')\; ,
\end{equation}\\
where
\begin{equation}
    f_{lm}^{l-1}(r') = \frac{\mathcal{G} \mathcal{M}_{\rm{pert}}}{r_{\rm{pert}}^2}\; 4\pi \sqrt{\frac{l}{2l+1}} \left(\frac{r'}{r_{\rm{pert}}}\right)^{l-1} Y^*_{lm}(\vartheta_{\rm{pert}}, \varphi_{\rm{pert}}) \; .
\end{equation}\\
In the above expressions, $\mathbf{Y}_{lm}^{l-1}(\vartheta, \varphi)$ stands for vectorial spherical harmonics and $Y_{lm}(\vartheta, \varphi)$ for scalar spherical harmonics \citep{varsalovic88}. The asterisk above $Y_{lm}$ indicates complex conjugation.\\

If the planet is perturbed by the primary, we have:

\begin{equation}
    r_{\rm{pert}} = r_A = a_{A} (1-e_{A} \cos{E_{A}})\; , \qquad \qquad \varphi_{\rm{pert}}=v_{\rm{A}} - \dot{\theta}_{A} t \; . \label{eq:coord_S}
\end{equation}\\
Under the assumption of zero obliquity, the colatitude is constant at $\vartheta_{\rm{pert}}=\pi/2$. The eccentric anomaly $E_{A}$ is calculated by iteratively solving Kepler's equation and the true anomaly $v_{A}$ is evaluated from its relation to the eccentric anomaly \citep[e.g.,][]{murray99}.

In the case of a perturbation caused by another secondary, designed previously as planet $B$, the coordinates are \citep[see also][]{hay19}:

\begin{equation}
    \begin{gathered}
        r_{\rm{pert}} = \sqrt{r_A^2 + r_B^2 - 2r_A r_B \cos{(v_B-v_A)}} \\
        \varphi_{\rm{pert}} = \alpha + v_{\rm{A}} - \dot{\theta}_{A} t \\
        \sin{\alpha} = \frac{r_{B}}{r_{\rm{pert}}} \sin{(v_A-v_B)} \qquad \cos{\alpha} = \frac{r_A - r_B \cos{(v_A-v_B)}}{r_{\rm{pert}}}
    \end{gathered}\label{eq:coord_B}
\end{equation}\\
For a coplanar configuration, the colatitude is, again, $\vartheta_{\rm{pert}}=\pi/2$.\\

The viscoelastic spherical shell deforms under the action of a volumetric tidal force. Specifically, the radial deformation field can be written as:

\begin{equation}
    u_{r}(r,\vartheta,\varphi) = \sum_{lm} u_{lm}(r) Y_{lm}(\vartheta, \varphi) \; ,
\end{equation}\\
where $u_{lm}(r)$ are coefficients of the spherical harmonic expansion at radial distance $r$ from the planet center. Surface deflections as well as the undulation of the core-mantle boundary (CMB) give rise to the disturbing force. Exterior to the planet, it reads as:

\begin{equation}
    \mathbf{f}_{\rm{dist}} (r,\vartheta,\varphi) = \sum_{lm} \left[\zeta_{lm}^{\rm{core}} \left(\frac{R_{\rm{core}}}{r}\right)^{l+2} + \zeta_{lm}^{\rm{surf}} \left(\frac{R_{\rm{surf}}}{r}\right)^{l+2}\right] \mathbf{Y}_{lm}^{l+1} (\vartheta, \varphi)\; ,
\end{equation}\\
where

\begin{equation}
    \zeta_{lm}^{\rm{core}}=4\pi\sqrt{\frac{l+1}{2l+1}}\,\mathcal{G} \Delta\rho\, u_{lm} (R_{\rm{core}})\qquad \mbox{and} \qquad \zeta_{lm}^{\rm{surf}}=4\pi\sqrt{\frac{l+1}{2l+1}}\,\mathcal{G}\,\rho\,u_{lm} (R_{\rm{surf}})\, , \label{eq:fintco}
\end{equation}\\
with $\rho$ being the spherical shell's density and $\Delta\rho$ the density contrast at CMB. As mentioned previously, for the sake of comparison with the analytical model, we set $\Delta\rho=0$ and, therefore, $\zeta_{lm}^{\rm{core}}=0$.

The torque acting on the deformed planet due to a body of mass $\mathcal{M}_2$ at coordinates $\mathbf{r}$ is then evaluated as:

\begin{equation}
\mathcal{T}=-\mathcal{M}_{\rm{2}}\; \mathbf{r} \times \mathbf{f}_{\rm{dist}}\, . \label{eq:torque}
\end{equation}\\
Depending on the considered component of the tidal torque, $m_{\rm{2}}$ can be either the mass of the primary or of the other secondary. The radial vector $\mathbf{r}$ is, accordingly, either evaluated from equation (\ref{eq:coord_S}) or (\ref{eq:coord_B}).

Finally, to obtain the evolution rates of the Keplerian elements, we insert the disturbing function $\mathbf{f}_{\rm{dist}}$ in the set of Gauss's planetary equations \citep[e.g.,][]{brouwer61,murray99}.

\section{Tidal potential in the general 3-D case} \label{sec:supplement}

In the main text, we derived the analytical formulae for the planet-planet (or moon-moon) tidal potential in a coplanar system. This appendix presents a generalization to a fully 3-D case, with possibly non-zero mutual inclinations and planetary obliquities. We also illustrate how the general case simplifies to a coplanar case, the tidal perturbation due to a central body \citep{kaula61,kaula64}, and the mutual gravitational perturbation of two point masses orbiting the same primary \citep{kaula62}.

\subsection{Derivation}

Consider a system consisting of a host star and two planets, denoted as planet $A$ and planet $B$. The reference coordinate system $\{xyz\}^{\rm{R}}$ is chosen such that its origin lies in the center of the host star, the $x^{\rm{R}}$ axis points towards the pericenter $\pi_{A}$ of planet $A$ and the orbit of planet $A$ lies in the $\{xy\}^{\rm{R}}$ plane. The semi-major axis and the eccentricity of planet $A$'s orbit are $a_{A}$ and $e_{A}$, respectively, and the Keplerian elements of planet $B$'s orbit are $\{a_{B}, e_{B}, i_{B}, \omega_{B}, \Omega_{B}, \sigma_{B}\}$.

In order to derive the disturbing function in the most general form, we further prescribe arbitrary orientation of planet $A$'s spin axis, defined by angles $\alpha$ (initial position of the meridian), $\beta$ (the obliquity) and $\gamma$ (position of the vernal point). Our goal is to provide a Fourier decomposition in space and time of the gravitational potential acting on the surface of planet $A$ due to the presence of planet $B$. The potential will be expressed as a function of Keplerian elements of the two planets. However, the spherical harmonic decomposition will be done with respect to the planet $A$'s surface, utilizing spherical planetocentric coordinates.\\

Throughout the derivation, we will work with three rotation matrices $\mathbf{R}^{\rm{orb}}$, $\mathbf{R}^{\rm{eqx}}$ and $\mathbf{R}^{\rm{plt}}$. First of them describes transformation from planet $B$'s orbit to the reference coordinate system,

\begin{equation}
\mathbf{R}^{\rm{orb}} =
  \begin{pmatrix}
  \cos{\omega_B}\,\cos{\Omega_B} - \sin{\omega_B}\,\sin{\Omega_B}\,\cos{i_B} & -\cos{\omega_B}\,\sin{\Omega_B} - \sin{\omega_B}\,\cos{\Omega_B}\,\cos{i_B} & \sin{\omega_B}\,\sin{i_B} \\
  \sin{\omega_B}\,\cos{\Omega_B} + \cos{\omega_B}\,\sin{\Omega_B}\,\cos{i_B} & -\sin{\omega_B}\,\sin{\Omega_B} + \cos{\omega_B}\,\cos{\Omega_B}\,\cos{i_B} & -\cos{\omega_B}\,\sin{i_B} \\
  \sin{\Omega_B}\,\sin{i_B} & \cos{\Omega_B}\,\sin{i_B} & \cos{i_B}
  \end{pmatrix} \label{eq:rorb}
\end{equation}\\

\noindent
the second one stands for the transformation from a planetocentric (``$A$-centric") coordinate system with axes parallel to $\{xyz\}^{\rm{R}}$ to a system whose $z$-axis aims at the direction of planet's rotational angular momentum and whose $x$-axis points towards the vernal equinox,

\begin{equation}
\mathbf{R}^{\rm{eqx}} =
  \begin{pmatrix}
  \cos{\gamma} & -\sin{\gamma} & 0 \\
  \cos{\beta}\,\sin{\gamma} & \cos{\beta}\,\cos{\gamma} & -\sin{\beta} \\
  \sin{\beta}\,\sin{\gamma} & \sin{\beta}\,\cos{\gamma} & \cos{\beta}
  \end{pmatrix}
\end{equation}\\

\noindent
and the last one is equal to $\mathbf{R}_{z}(\alpha)\,\mathbf{R}^{\rm{eqx}}$, where $\mathbf{R}_{z}(\alpha)$ symbolizes rotation by an angle $\alpha$ about the axis $z$, that is,

\begin{equation}
\mathbf{R}^{\rm{plt}} =
  \begin{pmatrix}
  \cos{\alpha}\,\cos{\gamma} - \sin{\alpha}\,\sin{\gamma}\,\cos{\beta} & -\cos{\alpha}\,\sin{\gamma} - \sin{\alpha}\,\cos{\gamma}\,\cos{\beta} & \sin{\alpha}\,\sin{\beta} \\
  \sin{\alpha}\,\cos{\gamma} + \cos{\alpha}\,\sin{\gamma}\,\cos{\beta} & -\sin{\alpha}\,\sin{\gamma} + \cos{\alpha}\,\cos{\gamma}\,\cos{\beta} & -\cos{\alpha}\,\sin{\beta} \\
  \sin{\gamma}\,\sin{\beta} & \cos{\gamma}\,\sin{\beta} & \cos{\beta}
  \end{pmatrix}
\end{equation}\\

With $r_{A}$ and $r_{B}$ being the instantaneous distances of the two planets from the host star and $v_A(t)$ and $v_{B}(t)$ denoting the respective true anomalies at time $t$, we first express the positions $\mathbf{r}^{\rm{R}}_{A}$ and $\mathbf{r}^{\rm{R}}_B$ of planets $A$ and $B$ in the reference coordinate system as

\begin{equation}
\mathbf{r}^{\rm{R}}_A (t) = r_A
\begin{pmatrix}
  \cos{v_A(t)} \\
  \sin{v_A(t)} \\
  0
\end{pmatrix} \label{eq:Ainer}
\end{equation}\\
and

\begin{equation}
\mathbf{r}^{\rm{R}}_B (t) = r_B\, \mathbf{R}^{\rm{orb}}
\begin{pmatrix}
  \cos{v_B(t)} \\
  \sin{v_B(t)} \\
  0
\end{pmatrix} \label{eq:Biner}
\end{equation}\\
\newline
Then, we transfer the origin of the coordinate system from the host star to the center of planet $A$ and transform the position vector of planet $B$ to the $A$-centric coordinate system $xyz$, where $z$ aims in the direction of planet $A$'s rotational angular momentum. Using a matrix notation, we are performing the following transformation

\begin{equation}
\mathbf{r}_{B} = \mathbf{R}^{\rm{plt}} \; \mathbf{r}\; ,\qquad \mathbf{r} = (\mathbf{r}_{B}^{\rm{R}}-\mathbf{r}_{A}^{\rm{R}})\; . \label{eq:Bplt}
\end{equation}\\
In the next step we are going to transform the position vector to spherical coordinates and find the longitude $\varphi$ and the colatitude $\vartheta$ of the "sub-$B$ point", as well as the instantaneous mutual distance $r=|\mathbf{r}|$. More specifically, we are seeking goniometric functions of the two coordinates. Tidal potential acting on infinitesimal volume at position $(r', \vartheta', \varphi')$ inside planet $A$ due to a perturber of mass $\mathcal{M}_{B}$ located at coordinates $(r, \vartheta, \varphi)$ can be expressed as

\begin{equation}
\begin{aligned}
\mathcal{U} &= \frac{\mathcal{G}\mathcal{M}_{B}}{r}\; \sum_{l=2}^{\infty}\left(\frac{r'}{r}\right)^{l}\sum_{m=0}^{l}(2-\delta_{m0})\frac{(l-m)!}{(l+m)!} \mathcal{P}_{lm}(\cos\vartheta) \mathcal{P}_{lm}(\cos\vartheta') \cos{[m(\varphi-\varphi')]} =\\[1.5ex]
&= \sum_{l=0}^{\infty} \frac{1}{r^{l+1}}\; \sum_{m=0}^{l}\mathcal{P}_{lm}(\cos\vartheta)\, \Big(\mathcal{A}_{lm}\cos{m\varphi}+\mathcal{B}_{lm}\sin{m\varphi}\Big)\; ,
\end{aligned} \label{eq:potential}
\end{equation}\\
where

\begin{align*}
\mathcal{A}_{lm} &= (2-\delta_{m0})\frac{(l-m)!}{(l+m)!}\; \mathcal{G}\mathcal{M}_{B}(r')^{l}\; \mathcal{P}_{lm}(\cos\vartheta')\; \cos{m\varphi'}\; , \\[1.5ex]
\mathcal{B}_{lm} &= (2-\delta_{m0})\frac{(l-m)!}{(l+m)!}\; \mathcal{G}\mathcal{M}_{B}(r')^{l}\; \mathcal{P}_{lm}(\cos\vartheta')\; \sin{m\varphi'}\; .
\end{align*}\\
and $\mathcal{P}_{lm}(x)$ denotes the associated Legendre polynomials of degree $l$ and order $m$. The latter can be defined by the following explicit formula.

\begin{equation}
\mathcal{P}_{lm}(x) = \frac{(1-x^2)^{\frac{m}{2}}}{2^l l!}\; \sum_{t=0}^{\lfloor\frac{l-m}{2}\rfloor} \frac{(2l-2t)!}{(l-m-2t)!} \binom{l}{t} (-1)^t x^{l-m-2t}\; . \label{eq:legendre}
\end{equation}\\

Let's begin with expressing the term $\cos\vartheta$, that is

\begin{equation}
\cos\vartheta=\frac{z_{B}}{r}=\frac{\mathbf{e}_{z}\,\mathbf{R}^{\rm{plt}}\,\mathbf{r}}{r}\; .
\label{eq:costhet}
\end{equation}\\

To derive the terms containing $\varphi$, we first introduce new variables, the right ascension $\rm{RA}$ and the sidereal time $\theta$, and then rewrite the longitude of sub-$B$ point as $\varphi=\rm{RA}-\theta$. It follows that

\begin{equation}
\begin{aligned}
\cos{m\varphi} = \cos{m(\rm{RA}-\alpha)}\,\cos{m(\alpha-\theta)} - \sin{m(\rm{RA}-\alpha)}\,\sin{m(\alpha-\theta)}\; , \\[1.5ex]
\sin{m\varphi} = \sin{m(\rm{RA}-\alpha)}\,\cos{m(\alpha-\theta)} + \cos{m(\rm{RA}-\alpha)}\,\sin{m(\alpha-\theta)}\, .
\end{aligned} \label{eq:mphi}
\end{equation}\\
Recalling trigonometric identities (3) and (4) of \citet{kaula61},

\begin{equation}
\begin{aligned}
\cos{mx} &= \Re\Bigg\{\sum_{s=0}^{m}\binom{m}{s} i^{s} \cos^{m-s}x \sin^{s}x \Bigg\}\; , \\[1.5ex]
\sin{mx} &= \Re\Bigg\{\sum_{s=0}^{m}\binom{m}{s} i^{s-1} \cos^{m-s}x \sin^{s}x \Bigg\}\; ,
\end{aligned}
\end{equation}\\
we can write 

\begin{equation}
\mathcal{A}_{lm}\cos{m\varphi}+\mathcal{B}_{lm}\sin{m\varphi} =\Re\Bigg\{\sum_{s=0}^{m}\binom{m}{s} i^{s} \cos^{m-s}({\rm{RA}}-\alpha) \sin^{s}({\rm{RA}}-\alpha) \big(\mathcal{A}_{lm}-i\mathcal{B}_{lm}\big) \exp\big[i\,m(\alpha-\theta)\big] \Bigg\}
\label{eq:abcossin}
\end{equation}\\
with
\begin{eqnarray}
\cos({\rm{RA}}-\alpha)=\frac{\mathbf{e}_{x}\,\mathbf{R}^{{\rm{eqx}}}\,\mathbf{r}}{r\sin{\vartheta}} \label{eq:cosra} \\[3ex]
\sin({\rm{RA}}-\alpha)=\frac{\mathbf{e}_{y}\,\mathbf{R}^{{\rm{eqx}}}\,\mathbf{r}}{r\sin{\vartheta}} \label{eq:sinra}
\end{eqnarray}\\
Finally, the instantaneous mutual distance $r$ is given by

\begin{equation}
r = \sqrt{x_{B}^2+y_{B}^2+z_{B}^2} = \sqrt{r_{A}^2+r_{B}^2-2\,r_{A}\,r_{B}\,\cos\mathcal{E}}\; ,
\label{eq:mutdist}
\end{equation}\\
where $\mathcal{E}$ symbolizes the angular distance between the two planets, as observed from the host star,

\begin{equation}
\cos\mathcal{E} = \cos{(v_{B}+\omega_{B})}\,\cos{(v_{A}-\Omega_{B})} + \cos{i_{B}}\,\sin{(v_{B}+\omega_{B})}\,\sin{(v_{A}-\Omega_{B})}\; . \nonumber
\end{equation}\\
Inserting (\ref{eq:costhet}) into (\ref{eq:legendre}) leads to

\begin{equation}
\mathcal{P}_{lm}(\cos\vartheta) = \frac{\sin^{m}\vartheta}{2^l l!}\; \sum_{t=0}^{\frac{l-m}{2}} \frac{(2l-2t)!}{(l-m-2t)!} \binom{l}{t} (-1)^t \cos^{l-m-2t}\vartheta \label{eq:legendrecos}
\end{equation}\\
where, upon applying multinomial expansion,

\begin{equation}
\begin{gathered}
\cos^{l-m-2t}\vartheta = \frac{1}{r^{l-m-2t}} \sum_{f_1=0}^{l-m-2t} \sum_{f_2=0}^{l-m-2t-f_1} \sum_{f_3=0}^{l-m-2t-f_1-f_2} \binom{l-m-2t}{f_1} \binom{l-m-2t-f_1}{f_2} \binom{l-m-2t-f_1-f_2}{f_3} \;\times \\[3ex]
\times\; r_B^{f_1+f_2+f_3} (-r_A)^{l-m-2t-f_1-f_2-f_3}\, (\sin\beta)^{l-m-2t}\, (\cos\beta)^{f_3}\, (\sin{i_B})^{f_3}\, (\cos{i_B})^{f_2} \;\times \\[3ex]
\times\;(\sin(\omega_B+\gamma))^{f_1}\, (\cos(\omega_B+\gamma))^{f_2}\, (\sin(\Omega_B+v_B))^{f_2+f_3}\, (\cos(\Omega_B+v_B))^{f_1}\, (\sin(v_A+\gamma))^{l-m-2t-f_1-f_2-f_3}\; .
\end{gathered} \label{eq:exp1}
\end{equation}\\

\noindent
Similarly, expressing the respective powers of (\ref{eq:cosra}) and (\ref{eq:sinra}) present in (\ref{eq:abcossin}) and using the multinomial expansion, we arrive at

\begin{equation}
\begin{gathered}
\Big[\cos({\rm{RA}}-\alpha)\Big]^{m-s} = \frac{1}{(r\sin\vartheta)^{m-s}} \sum_{c_1=0}^{m-s} \sum_{c_2=0}^{m-s-c_1} \binom{m-s}{c_1} \binom{m-s-c_1}{c_2} (-1)^{m-s-c_1}\, r_B^{c_1+c_2}\, r_A^{m-s-c_1-c_2}\; \times \\[3ex]
\times\; (\cos{i_B})^{c_2}\, (\sin(\omega_B+\gamma))^{c_2}\, (\cos(\omega_B+\gamma))^{c_1}\, (\sin(\Omega_B+v_B))^{c_2}\, (\cos(\Omega_B+v_B))^{c_1}\, (\cos(v_A+\gamma))^{m-s-c_1-c_2}
\end{gathered} \label{eq:exp2}
\end{equation}
and
\begin{equation}
\begin{gathered}
\Big[\sin({\rm{RA}}-\alpha)\Big]^{s} = \frac{1}{(r\sin\vartheta)^{s}} \sum_{d_1=0}^{s} \sum_{d_2=0}^{s-d_1} \sum_{d_3=0}^{s-d_1-d_2} \binom{s}{d_1} \binom{s-d_1}{d_2} \binom{s-d_1-d_2}{d_3} (-1)^{s-d_1-d_2}\; \times \\[3ex]
\times\;  r_B^{d_1+d_2+d_3}\, r_A^{s-d_1-d_2-d_3}\, (\sin\beta)^{d_3}\, (\cos\beta)^{s-d_3}\, (\sin{i_B})^{d_3}\, (\cos{i_B})^{d_2}\; \times \\[3ex]
\times\; (\sin(\omega_B+\gamma))^{d_1}\, (\cos(\omega_B+\gamma))^{d_2}\, (\sin(\Omega_B+v_B))^{d_2+d_3}\, (\cos(\Omega_B+v_B))^{d_1}\, (\cos(v_A+\gamma))^{s-d_1-d_2-d_3} \; .
\end{gathered} \label{eq:exp3}
\end{equation}\\
In the next step, we insert (\ref{eq:legendrecos}) with expansion (\ref{eq:exp1}) and (\ref{eq:abcossin}) with expansions (\ref{eq:exp2})-(\ref{eq:exp3}) into the general expression for tidal potential (\ref{eq:potential}). The terms $\sin^m{\vartheta}$, appearing in the numerator of (\ref{eq:legendrecos}) and in the denominator of (\ref{eq:abcossin}), cancel each other and we are left with a product of algebraic prefactors, physical parameters and goniometric functions in the form $\sin^a{x} \cos^b{x}$. Applying identity (5) of \citet{kaula61},

\begin{equation}
\sin^a{x} \cos^b{x} = \frac{(-i)^a}{2^{a+b}} \sum_{c=0}^{a} \sum_{d=0}^{b} \binom{a}{c}\binom{b}{d} (-1)^c \big[\cos(a+b-2c-2d)x+i\sin(a+b-2c-2d)x\big]\; , \label{eq:sinacosb}
\end{equation}\\
on the goniometric functions with arguments $(\omega_B+\gamma)$, $(\Omega_B+v_B)$ and $(v_A+\gamma)$, we may rewrite the tidal potential as

\begin{gather}
\mathcal{U} = \Re\Bigg\{\sum_{l=0}^{\infty} \sum_{m=0}^{l} \sum_{t=0}^{\lfloor\frac{l-m}{2}\rfloor} \frac{1}{r^{2l-2t+1}}\; \frac{i^{l-m}\, (-1)^{l+s-f_1-f_2-f_3-c_1-c_2-d_1-d_2}}{2^{2l-2t+c_1+c_2+d_1+d_2+f_1+f_2}}\; \frac{(2l-2t)!}{(l-m-2t)!\;l!}\; \binom{l}{t}\; \times \nonumber \\[3ex]
\times\; \sum_{|f|=l-m-2t} \binom{l-m-2t}{|f|}\; \sum_{s=0}^{m} \binom{m}{s} \sum_{|c|=m-s} \binom{m-s}{|c|} \sum_{|d|=s} \binom{s}{|d|}\; \times \nonumber\\[3ex]
\times\; r_B^{f_1+f_2+f_3+c_1+c_2+d_1+d_2+d_3}\; r_A^{l-2t-f_1-f_2-f_3-c_1-c_2-d_1-d_2-d_3} \; \times \nonumber\\[3ex]
\times\;(\sin\beta)^{l-m-2t-f_3+d_3}\; (\cos\beta)^{f_3+s-d_3}\; (\sin{i_B})^{f_3+d_3}\; (\cos{i_B})^{f_2+d_2+c_2} \;\times \nonumber\\[3ex]
\times\; \sum_{h_1=0}^{f_1+d_1+c_2} \sum_{h_2=0}^{f_2+d_2+c_1} \binom{f_1+d_1+c_2}{h_1}\; \binom{f_2+d_2+c_1}{h_2}\; \times \nonumber\\[3ex]
\times\; \sum_{k_1=0}^{f_2+f_3+c_2+d_2+d_3} \sum_{k_2=0}^{f_1+c_1+d_1} \binom{f_2+f_3+c_2+d_2+d_3}{k_1}\; \binom{f_1+c_1+d_1}{k_2}\; \times \nonumber\\[3ex]
\times\; \sum_{w_1=0}^{s-d_1-d_2-d_3+l-m-2t-f_1-f_2-f_3} \sum_{w_2=0}^{m-s-c_1-c_2} \binom{s-d_1-d_2-d_3+l-m-2t-f_1-f_2-f_3}{w_1}\; \binom{m-s-c_1-c_2}{w_2}\; \times \nonumber\\[3ex]
\times\; (-1)^{h_1+k_1+w_1}\; \big(\mathcal{A}_{lm}-i\mathcal{B}_{lm}\big)\; \exp\Big\{i\Big[(c_1+c_2+d_1+d_2+f_1+f_2-2h_1-2h_2)(\omega_B+\gamma) + \nonumber\\[3ex]
+ (c_1+c_2+d_1+d_2+d_3+f_1+f_2+f_3-2k_1-2k_2)(\Omega_B+v_B) + \nonumber\\[3ex]
+ (l-2t-c_1-c_2-d_1-d_2-f_1-f_2-f_3-2w_1-2w_2)(v_A+\gamma) + m(\alpha-\theta)\Big]\Big\} \Bigg\} \; .
\label{eq:potplt}
\end{gather}\\
The summation symbols and binomial coefficients with $|f|$, $|c|$, or $|d|$ are a shortened notation for the sets of summation symbols and binomial coefficients from equations (\ref{eq:exp1}-\ref{eq:exp3}). In the following, it will be practical to introduce a new set of indices, $\{p_1, p_2, p_3, p_4, p_5\}$, and to substitute them for the former ones,
\begin{align*}
d_2 &= p_1-d_1-c_1-c_2-f_1-f_2 \; , \\
d_3 &= p_2-p_1-f_3 \; , \\
w_2 &= p_3-w_1 \; , \\
k_2 &= p_4-k_1 \; , \\
h_2 &= p_5-h_1 \; ,
\end{align*}\\
together with changing the limits of summation over $d_1$, $h_1$, $k_1$, $w_1$, which is equivalent to only taking the non-zero terms of the corresponding expansion.\\

Next, we expand the $(2l-2t+1)$-th inverse power of the instantaneous mutual distance (\ref{eq:mutdist}) into a series of Gegenbauer polynomials with $r_{<}={\rm{min}}(r_A, r_B)$ and $r_{>}={\rm{max}}(r_A, r_B)$,

\begin{equation}
\frac{1}{r^{2l-2t+1}} = \frac{1}{r_{>}^{2l-2t+1}} \sum_{n=0}^{\infty} \left(\frac{r_{<}}{r_{>}}\right)^n \, \mathcal{C}_{n}^{\left(\frac{2l-2t+1}{2}\right)}(\cos\mathcal{E}) \; , \label{eq:gegexp}
\end{equation}\\
which can be defined explicitly as

\begin{equation}
\mathcal{C}_{n}^{(\lambda)}(x) = \sum_{u=0}^{\lfloor\frac{n}{2}\rfloor} (-1)^{u}\, \frac{\Gamma(n-u+\lambda)}{\Gamma(\lambda)\,u!(n-2u)!}\,(2x)^{n-2u} \; . \label{eq:gegenbauer}
\end{equation}\\
Symbol $\lfloor\frac{n}{2}\rfloor$ indicates the integer part of $\frac{n}{2}$. Substituting (\ref{eq:gegenbauer}) into (\ref{eq:gegexp}), where $\cos\mathcal{E}$ is given by (\ref{eq:mutdist}), and performing few algebraic steps identical to the previous expansion leads to

\begin{equation}
\begin{gathered}
\frac{1}{r^{2l-2t+1}} = \frac{1}{r_{>}^{2l-2t+1}} \sum_{n=0}^{\infty} \left(\frac{r_{<}}{r_{>}}\right)^n \, \sum_{u=0}^{\lfloor\frac{n}{2}\rfloor} (-1)^{u}\, \frac{\Gamma(n-u+\frac{2l-2t+1}{2})}{\Gamma(\frac{2l-2t+1}{2})\,u!(n-2u)!} \;\times \\[1.5ex]
\times\; \sum_{j=0}^{n-2u} \binom{n-2u}{j} \cos^j{i_B}\; \frac{(-1)^j}{2^{n-2u}} \sum_{r_1=0}^{j} \sum_{r_2=0}^{n-2u-j} \sum_{s_1=0}^{j} \sum_{s_2=0}^{n-2u-j} \binom{j}{r_1} \binom{n-2u-j}{r_2} \binom{j}{s_1} \binom{n-2u-j}{s_2} \; \times \\[1.5ex]
\; \times (-1)^{r_1+s_1} \exp\Big[i\,(n-2u-2r_1-2r_2)(v_B+\omega_B) + (n-2u-2s_1-2s_2)(v_A-\Omega_B)\Big]
\end{gathered} \label{eq:distgegen}
\end{equation}\\
Again, we will substitute indices in the argument of (\ref{eq:distgegen}) by $\{p_6,p_7\}$, so that
\begin{align*}
r_2 &= p_6-u-r_1 \; , \\
s_2 &= p_7-u-s_1 \; .
\end{align*}\\

Now, let's put the expressions (\ref{eq:potplt}) and (\ref{eq:distgegen}) together and introduce functions $\mathcal{D}^{{\rm{orb}}}_{I_1}(i_B)$, describing the 'orbital' portion of the tidal potential, and functions $\mathcal{F}^{{\rm{tilt}}}_{I_2}(i_B,\beta)$, in analogy with Kaula's functions of inclination:

\begin{equation}
\begin{gathered}
\mathcal{D}^{{\rm{orb}}}_{I_1}(i_B) = \sum_{u=0}^{\lfloor\frac{n}{2}\rfloor} (-1)^{u}\, \frac{\Gamma(n-u+\frac{2l-2t+1}{2})}{\Gamma(\frac{2l-2t+1}{2})\,u!(n-2u)!} \;\times \\[1.5ex]
\times\; \sum_{j=0}^{n-2u} \binom{n-2u}{j} \cos^j{i_B}\; \frac{(-1)^j}{2^{n-2u}} \sum_{r_1} \sum_{s_1} (-1)^{r_1+s_1} \binom{j}{r_1} \binom{n-2u-j}{p_6-u-r_1} \binom{j}{s_1} \binom{n-2u-j}{p_7-u-s_1}
\end{gathered} \label{eq:Dorb}
\end{equation}\\
and

\begin{equation}
\begin{gathered}
\mathcal{F}^{{\rm{tilt}}}_{I_2}(i_B,\beta) = \sum_{{\rm{indices}}} \frac{(-1)^{m+p_1+f_3+h_1+k_1+w_1}}{2^{2l-2t+p_1}} \frac{(2l-2t)!}{(l-m-2t)!\;l!} \;\times \\[3ex]
\times\; \binom{l}{t} \binom{m}{s} \binom{l-m-2t}{f_1} \binom{l-m-2t-f_1}{f_2} \binom{l-m-2t-f_1-f_2}{f_3} \binom{m-s}{c_1} \binom{m-s-c_1}{c_2}\; \times\\[3ex]
\times\; \binom{s}{d_1} \binom{s-d_1}{p_1-d_1-c_1-c_2-f_1-f_2} \binom{s-p_1+c_1+c_2+f_1+f_2}{p_2-p_1-f_3}  \binom{f_1+d_1+c_2}{h_1} \binom{p_1-f_1-d_1-c_2}{p_5-h_1}\;\times \\[3ex]
\times\;  \binom{p_2-c_1-d_1-f_1}{k_1} \binom{c_1+d_1+f_1}{p_4-k_1} \binom{l-m-2t+s-p_2+c_1+c_2}{w_1} \binom{m-s-c_1-c_2}{p_3-t-w_1} \; \times\\[3ex]
\times\; \big(\cos{i_B}\big)^{p_1-c_1-d_1-f_1} \big(\sin{i_B}\big)^{p_2-p_1} (\cos{\beta})^{2f_3-p_2+p_1+s} (\sin{\beta})^{l-m-2t+p_2-p_1-2f_3}
\end{gathered}
\end{equation}\\
\normalsize
with

\footnotesize
\begin{gather*}
\sum_{{\rm{indices}}} = \sum_{s=0}^{m} \sum_{f_1=0}^{l-m-2t} \sum_{f_2=0}^{l-m-2t-f_1} \sum_{f_3=0}^{l-m-2t-f_1-f_2} \sum_{c_1=0}^{m-s} \sum_{c_2=0}^{m-s-c_1} \sum_{d_1} \sum_{h_1} \sum_{k_1} \sum_{w_1}
\end{gather*}\\
\normalsize

\noindent
The potential is then rewritten as

\begin{equation}
\begin{gathered}
\mathcal{U} = \sum_{l=0}^{\infty} \sum_{m=0}^{l} \sum_{n=0}^{\infty} \sum_{t=0}^{\lfloor\frac{l-m}{2}\rfloor} \sum_{p_{\{1,\dots,7\}}} \frac{1}{r_{>}^{2l-2t+1}} \left(\frac{r_{<}}{r_{>}}\right)^n r_A^{l-2t-p_2} r_B^{p_2} \;\times \\[1.5ex] 
\times\; \mathcal{D}^{{\rm{orb}}}_{I_1}(i_B)\; \mathcal{F}^{{\rm{tilt}}}_{I_2}(i_B,\beta)\; \Bigg[
\begin{Bmatrix}
(-1)^{\frac{l-m}{2}}\mathcal{A}_{lm} \\
(-1)^{\frac{l-m-1}{2}}\mathcal{B}_{lm}
\end{Bmatrix}^{l-m\;{\rm{even}}}_{l-m\;{\rm{odd}}}
\cos\Big[(l+n-p_2-2p_3-2p_7)v_A \; + \\[1.5ex]
+ \;(n+p_2-2p_4-2p_6)v_B + (n+p_1-2p_5-2p_6)\omega_B + (-n+p_2-2p_4+2p_7)\Omega_B + \\[1.5ex]
+\; (l+p_1-p_2-2p_3-2p_5)\gamma+m(\alpha-\theta)\Big] \; + \\[1.5ex]
+\;
\begin{Bmatrix}
(-1)^{\frac{l-m}{2}}\mathcal{B}_{lm} \\
(-1)^{\frac{l-m+1}{2}}\mathcal{A}_{lm}
\end{Bmatrix}^{l-m\;{\rm{even}}}_{l-m\;{\rm{odd}}}
\sin\Big[(l+n-p_2-2p_3-2p_7)v_A \; + \\[1.5ex]
+ \;(n+p_2-2p_4-2p_6)v_B + (n+p_1-2p_5-2p_6)\omega_B + (-n+p_2-2p_4+2p_7)\Omega_B +\\[1.5ex]
+\; (l+p_1-p_2-2p_3-2p_5)\gamma+m(\alpha-\theta)\Big]\Bigg]\; ,
\end{gathered}
\end{equation}\\
where we used the abbreviation

\begin{equation}
\sum_{p_{\{1,\dots,7\}}} = \sum_{p_1=0}^{l-2t} \sum_{p_2=0}^{l-2t} \sum_{p_3=0}^{l-p_2} \sum_{p_4=0}^{p_2} \sum_{p_5=0}^{p_1} \sum_{p_6=0}^{n} \sum_{p_7=0}^{n}
\end{equation}\\

In the last step, we will expand the instantaneous distances from the host star and the true anomalies into series containing the respective semi-major axes $a_i$ and mean anomalies $M_i$. Let $X_{k}^{b,c}$ denote Hansen's coefficients \citep[e.g.,][]{murray99}. Then,

\begin{equation}
\left(\frac{r_i}{a_i}\right)^b e^{icv_i} = \sum_{k=-\infty}^{\infty} X_{k}^{b,c}(e_i) e^{ikM_i} \; . \label{eq:hansen}
\end{equation}\\
As in the main text, we distinguish two particular cases (under the assumption that the orbits are not crossing). The instantaneous distance of planet $B$ from the host star is either always larger than the distance of planet $A$,

\begin{equation}
a_B > a_A \frac{1+e_A}{1-e_B}\; ,
\end{equation}\\
or it is always smaller, hence
\begin{equation}
a_A > a_B \frac{1+e_B}{1-e_A}\; .
\end{equation}\\
In the first case, we may substitute:
\begin{equation*}
b_A = l+n-2t-p_2\; , \qquad \qquad b_B = -2l-n+2t+p_2-1\; ,
\end{equation*}
and in the second case:
\begin{equation*}
b_A = -l-n-p_2-1\; , \qquad \qquad b_B = n+p_2\; .
\end{equation*}
The other two indices are
\begin{align*}
c_A &= l+n-p_2-2p_3-2p_7 && c_B = n+p_2-2p_4-2p_6\\
k_A &= l+n-p_2-2p_3-2p_7+q_1 && k_B = n+p_2-2p_4-2p_6+q_2
\end{align*}\\
Introducing the functions of eccentricity $\mathcal{G}_{I_A}=X_{k_A}^{b_A,c_A}$ and $\mathcal{G}_{I_B}=X_{k_B}^{b_B,c_B}$, we may write down the final form of the tidal potential due to planet-planet tides as:

\begin{equation}
\begin{gathered}
\mathcal{U} = \sum_{l=0}^{\infty} \sum_{m=0}^{l} \sum_{n=0}^{\infty} \sum_{t=0}^{\lfloor\frac{l-m}{2}\rfloor} \sum_{p_{\{1,\dots,7\}}} \sum_{q_1=-\infty}^{\infty} \sum_{q_2=-\infty}^{\infty} a_A^{b_A}\,a_B^{b_B}\; \mathcal{D}^{{\rm{orb}}}_{I_1}(i_B)\; \mathcal{F}^{{\rm{tilt}}}_{I_2}(i_B,\beta)\; \mathcal{G}_{I_A}(e_A)\; \mathcal{G}_{I_B}(e_B) \;\times \\[1.5ex] 
\times\; \Bigg[
\begin{Bmatrix}
(-1)^{\frac{l-m}{2}}\mathcal{A}_{lm} \\
(-1)^{\frac{l-m-1}{2}}\mathcal{B}_{lm}
\end{Bmatrix}^{l-m\;{\rm{even}}}_{l-m\;{\rm{odd}}}
\cos\Big[(l+n-p_2-2p_3-2p_7+q_1)M_A \; + \\[1.5ex]
+ \;(n+p_2-2p_4-2p_6+q_2)M_B + (n+p_1-2p_5-2p_6)\omega_B + (-n+p_2-2p_4+2p_7)\Omega_B + \\[1.5ex]
+\; (l+p_1-p_2-2p_3-2p_5)\gamma+m(\alpha-\theta)\Big] \; + \\[1.5ex]
+\;
\begin{Bmatrix}
(-1)^{\frac{l-m}{2}}\mathcal{B}_{lm} \\
(-1)^{\frac{l-m+1}{2}}\mathcal{A}_{lm}
\end{Bmatrix}^{l-m\;{\rm{even}}}_{l-m\;{\rm{odd}}}
\sin\Big[(l+n-p_2-2p_3-2p_7+q_1)M_A \; + \\[1.5ex]
+ \;(n+p_2-2p_4-2p_6+q_2)M_B + (n+p_1-2p_5-2p_6)\omega_B + (-n+p_2-2p_4+2p_7)\Omega_B +\\[1.5ex]
+\; (l+p_1-p_2-2p_3-2p_5)\gamma+m(\alpha-\theta)\Big]\Bigg]\; .
\end{gathered} \label{eq:potfin}
\end{equation}
The corresponding tidal frequencies can be evaluated as in the main text.

\subsection{Special configurations}

\subsubsection{Coplanar orbits: $i_B=0$, $\Omega_B+\omega_B=\varpi_B$}

If the planetary orbits are coplanar (but the obliquity of planet $A$ is still generally non-zero), the angles $\omega_B$ and $\Omega_B$ lie in the same plane and they can be substituted by the longitude of pericenter $\varpi_B=\Omega_B+\omega_B$. This leads to the simplification of the rotation matrix $\mathbf{R}^{\rm{orb}}$, expression (\ref{eq:rorb}), and of equations (\ref{eq:exp1}-\ref{eq:exp3}). Additionally, equation (\ref{eq:distgegen}) becomes

\begin{equation}
\begin{gathered}
\frac{1}{r^{2l-2t+1}} = \frac{1}{r_{>}^{2l-2t+1}} \sum_{n=0}^{\infty} \left(\frac{r_{<}}{r_{>}}\right)^n \, \sum_{u=0}^{\lfloor\frac{n}{2}\rfloor} (-1)^{u}\, \frac{\Gamma(n-u+\frac{2l-2t+1}{2})}{\Gamma(\frac{2l-2t+1}{2})\,u!(n-2u)!}\;\times \\[1.5ex]
\times\; \sum_{j=0}^{n-2u} \binom{n-2u}{j} \exp\Big[i(n-2u-2j)(v_B-v_A+\varpi_B)\Big]
\end{gathered} \label{eq:spec1gegen}
\end{equation}\\

The derivation of tidal potential from (\ref{eq:potential}) with the aforementioned simplifications yields:

\begin{equation}
\begin{gathered}
\mathcal{U} = \sum_{l=0}^{\infty} \sum_{n=0}^{\infty} \sum_{m=0}^{l} \sum_{t=0}^{\lfloor\frac{l-m}{2}\rfloor} \sum_{p_1=0}^{l-2t} \sum_{p_2=0}^{p_1} \sum_{p_3=0}^{l-2t-p_1} \sum_{p_4=0}^{n} \sum_{q_1=-\infty}^{\infty} \sum_{q_2=-\infty}^{\infty} a_A^{b_A}\,a_B^{b_B}\; \mathcal{F}_{I}^{{\rm{copl}}}(\beta)\; \mathcal{G}_{I_A}(e_A)\; \mathcal{G}_{I_B}(e_B) \;\times\\[1.5ex]
\times\;\Bigg[
\begin{Bmatrix}
(-1)^{\frac{l-m}{2}}\mathcal{A}_{lm} \\
(-1)^{\frac{l-m-1}{2}}\mathcal{B}_{lm}
\end{Bmatrix}^{l-m\;{\rm{even}}}_{l-m\;{\rm{odd}}}
\cos\Big[(l-n-p_1-2t-2p_3+2p_4+q_1)M_A \; + \\[1.5ex]
+ \;(n+p_1-2p_2-2p_4+q_2)M_B + (n+p_1-2p_2-2p_4)\varpi_B + (l-2t-2p_2-2p_3)\gamma+m(\alpha-\theta)\Big] \; + \\[1.5ex]
+\;
\begin{Bmatrix}
(-1)^{\frac{l-m}{2}}\mathcal{B}_{lm} \\
(-1)^{\frac{l-m+1}{2}}\mathcal{A}_{lm}
\end{Bmatrix}^{l-m\;{\rm{even}}}_{l-m\;{\rm{odd}}}
\sin\Big[(l-n-p_1-2t-2p_3+2p_4+q_1)M_A \; + \\[1.5ex]
+ \;(n+p_1-2p_2-2p_4+q_2)M_B + (n+p_1-2p_2-2p_4)\varpi_B + (l-2t-2p_2-2p_3)\gamma+m(\alpha-\theta)\Big]\Bigg]
\end{gathered} \label{eq:spec1potfin}
\end{equation}\\
with obliquity functions

\begin{equation}
\begin{gathered}
\mathcal{F}_{I}^{{\rm{copl}}}(\beta) = \sum_{u} (-1)^{u}\, \frac{\Gamma(n-u+\frac{2l-2t+1}{2})}{\Gamma(\frac{2l-2t+1}{2})\,u!(n-2u)!}\; \binom{l}{t} \binom{n-2u}{p_4-u}\; \frac{(2l-2t)!}{(l-m-2t)!\;l!} \frac{1}{2^{2l-2t}} \; \times \\[1.5ex]
\times \; \sum_{s=0}^{m} \binom{m}{s} \sum_{c_1=0}^{m-s} \sum_{d_1} \binom{l-m-2t}{p_1-c_1-d_1} \binom{m-s}{c_1} \binom{s}{d_1} (-1)^{m+p_1} \; \times \\[1.5ex]
\times \;\sum_{h_1} \sum_{k_1} \binom{p_1-c_1}{h_1} \binom{c_1}{p_2-h_1} \binom{l-m-2t+s-p_1+c_1}{k_1} \binom{m-s-c_1}{p_3-k_1} (-1)^{h_1+k_1} \; \times \\[1.5ex]
\times \; (\sin\beta)^{l-m-2t} (\cos\beta)^s
\end{gathered} \label{eq:spec1f}
\end{equation}\\
and the following indices in (\ref{eq:hansen}):

\begin{align*}
c_A &= l-n-p_1-2t-2p_3-2p_4 && c_B = n+p_1-2p_2-2p_4\\
k_A &= l-n-p_1-2t-2p_3-2p_4+q_1 && k_B = n+p_1-2p_2-2p_4+q_2
\end{align*}\\
Coefficients $b_A$ and $b_B$ are the same as in the previous section, except that $p_2$ has been replaced by $p_1$. Given explicitly,
\begin{align*}
&b_A = l+n-2t-p_1 && b_B = -2l-n+2t+p_1-1 && \mbox{for} \qquad a_B > a_A \frac{1+e_A}{1-e_B}
\end{align*}
and
\begin{align*}
&b_A = -l-n-p_1-1 && b_B = n+p_1 && \mbox{for} \qquad a_A > a_B \frac{1+e_B}{1-e_A}
\end{align*}

\subsubsection{Coplanar orbits, aligned spin axis: $\beta=0$, $i_B=0$, $\Omega_B+\omega_B=\varpi_B$}
Results of the previous subsection can be further simplified by assuming that the rotational axis of planet $A$ is perpendicular to the orbit, as is done in the main text. Then, the only non-zero terms are those satisfying $l-m-2t=0$ and we can also set $\gamma=0$. The resulting tidal potential reads as

\begin{equation}
\begin{gathered}
\mathcal{U} = \sum_{l=0}^{\infty} \sum_{n=0}^{\infty} \sum_{m=0}^{l} \sum_{p_1=0}^{m} \sum_{p_2=0}^{m-p_1} \sum_{p_3=0}^{p_1} \sum_{p_4=0}^{n} \sum_{q_1=-\infty}^{\infty} \sum_{q_2=-\infty}^{\infty} a_A^{b_A}\,a_B^{b_B}\; \mathcal{F}_{I}^{{\rm{align}}}\; \mathcal{G}_{I_A}(e_A)\; \mathcal{G}_{I_B}(e_B) \;\times\\[1.5ex]
\times\;\Bigg[
\begin{Bmatrix}
\mathcal{A}_{lm} \\
-\mathcal{B}_{lm}
\end{Bmatrix}^{l-m\;{\rm{even}}}_{l-m\;{\rm{odd}}}
\cos\Big[(m-p_1-2p_2-n+2p_4+q_1)M_A \; + \\[1.5ex]
+ \;(p_1-2p_3+n-2p_4+q_2)M_B + (p_1-2p_3+n-2p_4)\varpi_B + m(\alpha-\theta)\Big] \; + \\[1.5ex]
+\;
\begin{Bmatrix}
\mathcal{B}_{lm} \\
\mathcal{A}_{lm}
\end{Bmatrix}^{l-m\;{\rm{even}}}_{l-m\;{\rm{odd}}}
\sin\Big[(m-p_1-2p_2-n+2p_4+q_1)M_A \; + \\[1.5ex]
+ \;(p_1-2p_3+n-2p_4+q_2)M_B + (p_1-2p_3+n-2p_4)\varpi_B + m(\alpha-\theta)\Big]\Bigg]
\end{gathered} \label{eq:specXpotfin}
\end{equation}\\
and the algebraic prefactors have the following form

\begin{equation}
\begin{gathered}
\mathcal{F}_{I}^{{\rm{align}}} = \frac{1}{2^{l+m}}\; \frac{(l+m)!}{l!}\; \binom{l}{\lfloor\frac{l-m}{2}\rfloor} (-1)^{\lfloor\frac{l-m}{2}\rfloor} \sum_{s=0}^{m} \binom{m}{s} \sum_{d_1} \binom{m-s}{p_1-d_1} (-1)^{m-p_1} \; \times \\[1.5ex]
\times\; \sum_{h_1} \sum_{k_1} (-1)^{h_1+k_1} \binom{d_1}{h_1} \binom{p_1-d_1}{p_3-h_1} \binom{s-d_1}{k_1} \binom{m-s-p_1+d_1}{p_2-k_1} \; \times \\[1.5ex]
\times\; \sum_{u} (-1)^{u} \frac{\Gamma\left(n-u+\frac{l+m+1}{2}\right)}{\Gamma\left(\frac{l+m+1}{2}\right)\,u!\,(n-2u)!}\; \binom{n-2u}{p_4-u} \; .
\end{gathered}
\end{equation}\\

\subsubsection{Perturbation due to a central body: $a_B=0$}
The only nonzero terms in the limit of a single satellite orbiting a single central body are those with $b_B=n+p_1=0$. Setting $n=0$ and $p_1=0$ in (\ref{eq:spec1potfin}) and (\ref{eq:spec1f}) leads to

\begin{equation}
\begin{gathered}
\mathcal{U} = \sum_{l=0}^{\infty} \sum_{m=0}^{l} \sum_{t=0}^{\lfloor\frac{l-m}{2}\rfloor} \sum_{p_3=0}^{l-2t} \sum_{q_1=-\infty}^{\infty} a_A^{-l-1}\; \mathcal{F}_{I}^{{\rm{central}}}(\beta)\; \mathcal{G}_{I_A}(e_A)\; \times\\[1.5ex]
\times\;\Bigg[
\begin{Bmatrix}
(-1)^{\frac{l-m}{2}}\mathcal{A}_{lm} \\
(-1)^{\frac{l-m-1}{2}}\mathcal{B}_{lm}
\end{Bmatrix}^{l-m\;{\rm{even}}}_{l-m\;{\rm{odd}}}
\cos\Big[(l-2t-2p_3+q_1)M_A + (l-2t-2p_3)\gamma+m(\alpha-\theta)\Big] \; + \\[1.5ex]
+\;
\begin{Bmatrix}
(-1)^{\frac{l-m}{2}}\mathcal{B}_{lm} \\
(-1)^{\frac{l-m+1}{2}}\mathcal{A}_{lm}
\end{Bmatrix}^{l-m\;{\rm{even}}}_{l-m\;{\rm{odd}}}
\sin\Big[(l-2t-2p_3+q_1)M_A + (l-2t-2p_3)\gamma+m(\alpha-\theta)\Big]\Bigg]
\end{gathered} \label{eq:spec2potfin}
\end{equation}\\
with

\begin{equation}
\begin{gathered}
\mathcal{F}_{I}^{{\rm{central}}}(\beta) = \frac{1}{2^{2l-2t}}\; \frac{(2l-2t)!}{(l-m-2t)!\;l!}\; \binom{l}{t} \sum_{s=0}^{m} \binom{m}{s} (-1)^{m} \sum_{k_1} \binom{l-m-2t+s}{k_1}\; \times \\[1.5ex]
\times \; \binom{m-s}{p_3-k_1} (-1)^{k_1} (\sin\beta)^{l-m-2t} (\cos\beta)^s \; ,
\end{gathered} \label{eq:spec2f}
\end{equation}\\
If we substitute $p_3=p-t$ and change the notation $k_1\rightarrow c$, equations (\ref{eq:spec2potfin}) and (\ref{eq:spec2f}) transform into the equation (27) of \citet{kaula61} except for a factor $(-1)^l$, which is a consequence of our definition of the instantaneous mutual distance in equation (\ref{eq:Bplt}).

\subsubsection{Two point masses: $l=0$}
Finally, for the sake of completeness, we derive the special case of two gravitationally interacting point-mass secondaries. Setting $l=0$ leaves us only with the terms associated with the power series expansion of mutual distance, that is

\begin{equation}
\begin{gathered}
\mathcal{U} = \mathcal{G} m_B\, \sum_{n=0}^{\infty}  \frac{a_{<}^{n}}{a_{>}^{n+1}} \sum_{p_6=0}^{n} \sum_{p_7=0}^{n} \mathcal{D}^{{\rm{orb}}}_{I_1}(i_B) \sum_{q_1=-\infty}^{\infty} \sum_{q_2=-\infty}^{\infty}  \mathcal{G}_{I_A}(e_A)\; \mathcal{G}_{I_B}(e_B) \;\times \\[1.5ex] 
\times\;\cos\Big[(n-2p_7+q_1)M_A + (n-2p_6+q_2)M_B + (n-2p_6)\omega_B + (-n+2p_7)\Omega_B \Big]
\end{gathered} \label{eq:spec3potfin}
\end{equation}\\
with $\mathcal{D}^{{\rm{orb}}}_{I_1}(i_B)$ given by equation (\ref{eq:Dorb})\footnote{Note that in the case of $l=0$ and $t=0$, Gegenbauer polynomials reduce to Legendre polynomials.},

\begin{gather*}
\mathcal{D}^{{\rm{orb}}}_{I_1}(i_B) = \frac{1}{2^n}\sum_{u=0}^{\lfloor\frac{n}{2}\rfloor} (-1)^{u}\, \binom{2n-2u}{n} \binom{n}{u} \;\times \\[1.5ex]
\times\; \sum_{j=0}^{n-2u} \binom{n-2u}{j} \cos^j{i_B}\; \frac{(-1)^j}{2^{2n-4u}} \sum_{r_1} \sum_{s_1} (-1)^{r_1+s_1} \binom{j}{r_1} \binom{n-2u-j}{p_6-u-r_1} \binom{j}{s_1} \binom{n-2u-j}{p_7-u-s_1}\; .
\label{eq:spec3Dorb}
\end{gather*}\\

Note that function (\ref{eq:spec3Dorb}) is \textit{not} equivalent to the product of Kaula's inclination functions $\mathcal{F}_{nmp}(i_A)\,\mathcal{F}_{nmh}(i_B)$, used in his expansion of the luni-solar disturbing function \citep{kaula62}. Our expression for the disturbing potential, equation (\ref{eq:spec3potfin}), uses relative Keplerian elements of planet $B$ with respect to planet $A$, whilst \citet{kaula62} works with the Keplerian elements of both bodies referred to an independent coordinate system. Nevertheless, when summed over all indices, expression (\ref{eq:spec3potfin}) is equal to expression (10) of \citet{kaula62}.

\bibliography{sample701}{}
\bibliographystyle{aasjournalv7}



\end{document}